\documentclass[preprint,12pt]{elsarticle}

\usepackage[normalem]{ulem}  
\usepackage{color} 

\renewcommand\sout{\bgroup \color{red} \ULdepth=-.5ex \ULset}

\usepackage{amssymb,amsmath}

\journal{Nuclear Physics A}

\begin{document}

\begin{frontmatter}



\title{Chiral SU(3) theory of antikaon-nucleon interactions with improved threshold constraints}


\author[a,b]{Yoichi Ikeda}
\author[a]{Tetsuo Hyodo}
\author[c]{Wolfram Weise}
\address[a]{Department of Physics, Tokyo Institute of Technology, Meguro 152-8551, Japan}
\address[b]{RIKEN Nishina Center, 2-1, Hirosawa, Wako, Saitama 351-0198, Japan}
\address[c]{Physik-Department, Technische Universit\"at M\"unchen, D-85747 Garching, Germany}

\begin{abstract}
$\bar{K}$-nucleon interactions are investigated in the framework of coupled-channels dynamics based on the next-to-leading order chiral SU(3) meson-baryon effective Lagrangian. 
A recent determination of the 1s shift and width of kaonic hydrogen enables us to set accurate constraints on the coupled-channels meson-baryon amplitudes in the strangeness $S=-1$ sector. Theoretical uncertainties in the subthreshold extrapolation of the coupled-channels amplitudes are discussed. Using this framework, we give predictions for $K^-$-neutron interactions and for the spectrum of the $\Lambda(1405)$ resonance. A simplified, effective three-channel model using leading order chiral SU(3) meson-baryon interactions is also constructed for convenient application in $\bar{K}$-nuclear few-body calculations.
\end{abstract}


\end{frontmatter}



\section{Introduction}

An important and challenging theme in strangeness nuclear physics is the dynamics of antikaons interacting with nucleons and nuclei. The $\bar{K}N$ interaction at low energy is strongly attractive and generates the $\Lambda(1405)$ resonance as a quasi-bound state embedded in the $\pi\Sigma$ continuum below $\bar{K}N$ threshold~\cite{Dalitz}. One therefore expects that interesting phenomena will also take place when the antikaon is injected or stopped in nuclear systems. Much work has recently been devoted to investigations of  $\bar{K}$ few-nucleon systems~\cite{Kfewbody1, Kfewbody2,Kfewbody3} and of possible bound states of a $\bar{K}$ in heavier nuclei
~\cite{Kmedium}. So far, however, the uncertainties in the subthreshold extrapolation of $\bar{K}N$ interactions, apart from experimental ambiguities, have prohibited firm and consistent conclusions.

The basis for these studies is the $\bar{K}N$ two-body interaction. In addition to the strong attraction in the elastic
$\bar{K}N$ channel, the other prominent feature is the almost equally strong coupling in the transition amplitude 
for $\bar{K}N \leftrightarrow \pi\Sigma$. Such strong coupled-channels dynamics is well treated by a unitary approach starting from chiral SU(3)$_{R}\times$SU(3)$_{L}$ effective field theory~\cite{Kaiser1995, Oller2001, Borasoy2005, Borasoy2006, HJ2011}. 

In earlier work, the data base for low-energy $\bar{K}N$ interactions came from scattering experiments performed from the 1960s to the 1980s. However, the $K^{-}p$ total cross section data in elastic and inelastic channels and threshold branching ratios did not constrain the scattering amplitude sufficiently well, so that the extrapolation of the $\bar{K}N$ amplitude to the subthreshold energy region still suffered from large uncertainties~\cite{Borasoy2005, Borasoy2006}.

Information about $\bar{K}N$ threshold physics comes primarily from measurements of the energy shift, $\Delta E$, and width, $\Gamma$, of the $1s$ state in kaonic hydrogen.  From these measurements, the real and imaginary parts of the $K^{-}p$ scattering length, $a(K^-p)$, can be deduced applying the improved Deser-Trueman formula~\cite{MRR2004} (see also Ref.~\cite{CS2007} for a precise calculation of the atomic level shifts). A precise determination of  $a(K^-p)$ is crucial for a reliable subthreshold extrapolation of the corresponding scattering amplitudes. Several previous kaonic hydrogen measurements~\cite{Iwasaki1997,Beer2005} extracted $\Delta E$ and $\Gamma$, but with still large uncertainties. Moreover, a possible quantitative inconsistency between the DEAR measurements \cite{Beer2005} and the scattering data was pointed out in several theoretical papers~\cite{Borasoy2005, Borasoy2006, MRR2004}. The situation has now improved significantly with the advent of the new SIDDHARTA measurements of kaonic hydrogen~\cite{Bazzi2011}. Not only are these data far more precise than the previous ones, but as reported in very recent theoretical work \cite{Ikeda:2011pi}, they are also found to be fully consistent with the existing scattering data. This new result enables us to update the theoretical description of the $\bar{K}N$ interaction and to reduce the uncertainties in the subthreshold extrapolation of the $\bar{K}N$ amplitude. At the same time it is possible to predict improved values for the $K^-$ neutron scattering length, $a(K^-n)$, and to set constraints for the $K^-$-deuteron scattering length \cite{DM2011}.

In this paper, we present a further extended study of the $\bar{K}N$-$\pi\Sigma$ interaction using the chiral coupled-channels framework with constraints from the SIDDHARTA measurement together with the existing scattering data. The interaction kernel is constructed based on input from the chiral SU(3) meson-baryon effective Lagrangian. In addition to the results already reported in our previous letter~\cite{Ikeda:2011pi}, we discuss implications for the structure of the $\Lambda(1405)$ and $K^{-}n$ scattering.

\section{Chiral SU(3) dynamics at next-to-leading order}
\subsection{Chiral effective Lagrangian}

Our starting point is the chiral SU(3)$_{R}\times$SU(3)$_{L}$ meson-baryon effective Lagrangian 
at next-to-leading order (NLO):
\begin{eqnarray} 
{\cal L}_{\textit{eff}}({\cal B, U}) = 
{\cal L}_M({\cal U}) + {\cal L}_{MB}^{(1)}({\cal B,U}) 
+  {\cal L}_{MB}^{(2)}({\cal B,U})~~,
\end{eqnarray}
where ${\cal L}_M({\cal U})$ with ${\cal U} = u^2 = \exp[\textrm{i}\sqrt{2}\,\Phi/f]$ is the non-linear chiral meson Lagrangian incorporating the octet of pseudoscalar Nambu-Goldstone bosons $(\pi, K, \bar{K}, \eta)$ in the standard $3\times 3$ matrix representation $\Phi$. At this stage $f \simeq 86$ MeV is the pseudoscalar decay constant in the chiral limit. 

The baryon octet fields $(N, \Lambda, \Sigma, \Xi)$ are collected in the $3\times 3$ matrix ${\cal B}$. The pseudoscalar meson octet couples to the baryons through the mesonic vector current
\begin{eqnarray} 
v^\mu =  \frac{1}{2\textrm{i}}(u^\dagger\partial^\mu u + u\,\partial^\mu u^\dagger)~~,
\end{eqnarray}
and the axial vector current
\begin{eqnarray} 
a^\mu = \frac{1}{2\textrm{i}}(u^\dagger\partial^\mu u - u\,\partial^\mu u^\dagger) \equiv -\frac{1}{2}u^\mu~~.
\end{eqnarray}
The most general form of the meson-baryon interaction Lagrangian at leading order ${\cal O}(p)$ in the chiral expansion is given by
\begin{eqnarray} 
{\cal L}_{MB}^{(1)} = 
\mathrm{Tr}\Big(\bar{\cal B}(\textrm{i}\gamma_\mu {\cal D}^{\mu} - M_0){\cal B} 
- D~ \bar{\cal B}\,\gamma_\mu \gamma_5 \{a^\mu,{\cal B}\}  
- F ~\bar{\cal B}\,\gamma_\mu \gamma_5 [a^\mu,{\cal B}]\Big)~~,
\label{eq1}
\end{eqnarray}
with the chiral covariant derivative ${\cal D}^{\mu}{\cal B} = \partial^\mu {\cal B} + \textrm{i}[v^\mu,{\cal B}]$. Here $D$ and $F$ are the low energy constants of the axial vector couplings and $M_0$ is the baryon octet mass in the chiral limit. The vector current coupling to the baryons involves even numbers of pseudoscalar mesons, while the axial vector vertices involve odd numbers of mesons. At next-to-leading order, ${\cal O}(p^2)$, the Lagrangian introduces several low-energy constants ($b_i$ and $d_j$) as
\begin{align} 
{\cal L}_{MB}^{(2)} =& 
b_0\, \mathrm{Tr}\big(\bar{\cal B}\,{\cal B}\big)\,
\mathrm{Tr}\big(\chi_+\big)  
+ b_D \,\mathrm{Tr}\big(\bar{\cal B}\{\chi_+ , {\cal B}\}\big) + b_F\, 
\mathrm{Tr}\big(\bar{\cal B}[\chi_+ , {\cal B}]\big)\nonumber \\ 
&+ d_1\,\mathrm{Tr}\big(\bar{\cal B}\,\{u_\mu, [u^\mu, {\cal B}]\}\big) 
+ d_2\,\mathrm{Tr}\big(\bar{\cal B}\,[u_\mu, [u^\mu, {\cal B}]]\big) \nonumber\\
&+ d_3\,\mathrm{Tr}\big(\bar{\cal B}\,u_\mu\big)\, \mathrm{Tr}\big({\cal B}\,u^\mu\big) 
+ d_4\,\mathrm{Tr}\big(\bar{\cal B}\,{\cal B}\big)\,\mathrm{Tr}\big(u_\mu\, u^\mu\big)~~, 
\label{eq2}
\end{align}
where
\begin{eqnarray}
\chi_+ =  -{2\,\langle 0| \bar qq |0\rangle\over f^2}\,\big(u{\cal M}u + u^\dagger {\cal M}u^\dagger\big)
\end{eqnarray}
 is the explicit symmetry breaking term with the chiral condensate $\langle 0| \bar qq |0\rangle$ and the quark mass matrix ${\cal M} = \text{diag}(m_u, m_d, m_s)$. 
 
At tree level in chiral perturbation theory (ChPT), the low-energy constants $b_0$, $b_D$ and $b_F$ are related to the baryon octet masses. The present analysis goes systematically beyond tree level and utilizes chiral SU(3) effective field theory to determine the interaction kernel of the coupled-channels scattering equations in which this matrix kernel is iterated to all orders. Therefore the low energy constants need not be identical to those in ChPT once the renormalization of the one-particle irreducible graphs is properly taken into account.

%
\begin{figure}
\begin{center}
\includegraphics*[totalheight=1.6cm]{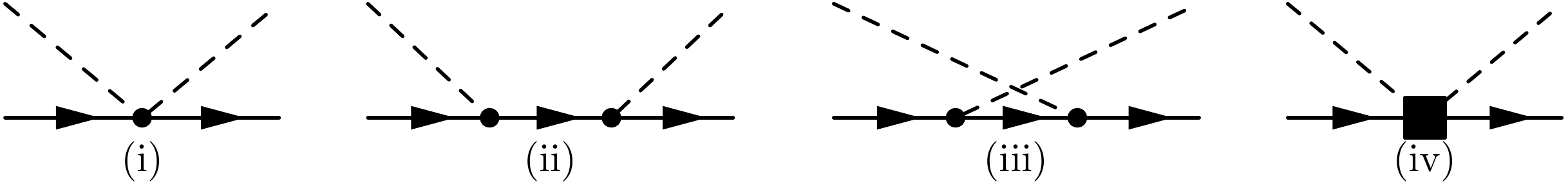}
\caption{Feynman diagrams for the meson-baryon interaction: Tomozawa-Weinberg term (i), direct and crossed Born terms (ii) and (iii), and NLO terms (iv). Dashed (solid) lines represent the pseudoscalar octet mesons (octet baryons). }
\label{Fig1}
\end{center}
\end{figure}
%

\subsection{Meson-baryon interactions}
Consider meson-baryon scattering using the chiral Lagrangians (\ref{eq1}) and (\ref{eq2}). In the SU(3) sector, several meson-baryon channels are coupled in sectors with given quantum numbers. The interaction matrix elements are  written as $\hat{V}_{ij}\equiv \langle i|\hat{V}|j\rangle$ with final and initial channel indices, $i$ and $j$. The strangeness $S=-1$ and charge $Q=0$ sector involves ten channels labeled by the indices $i = 1, \dots, 10$ in the order $K^-p$, $\bar{K}^0n$,  $\pi^0\Lambda$,  $\pi^0\Sigma^0$,  $\pi^+\Sigma^-$, $\pi^-\Sigma^+$,  $\eta\Lambda$,  $\eta\Sigma^0$,  $K^+\Xi^-$, $K^0\Xi^0$. The interaction $\hat{V}_{ij}$ is a function of the meson-baryon center-of-mass energy $\sqrt{s}$, the scattering angles $\Omega = \{\theta,\varphi\}$, and the baryon spin degrees of freedom $\sigma_{i}$ and $\sigma_{j}$. Projecting onto $s$-wave, the interactions depend only on $\sqrt{s}$:
\begin{eqnarray}
V_{ij}(\sqrt{s}) = \frac{1}{8\pi}\sum_\sigma\int d\Omega\, \hat{V}_{ij}(\sqrt{s}, \Omega, \sigma_i, \sigma_j)~~, 
\end{eqnarray}
where the spin average over $\sigma_i = \sigma_j = \pm 1/2$ is taken. 

The covariant derivative in Eq.~(\ref{eq1}) generates the Tomozawa-Weinberg contact term shown in Fig.~\ref{Fig1}(i). This is the leading-order contribution to pseudoscalar meson-baryon scattering, given as\footnote{The normalization convention used here is the same as in Ref.~\cite{Borasoy2005}, with dimensionless $V_{ij}$. It differs from the one used in Ref.~\cite{Oller2001,HJ2011} by a factor $\sqrt{M_i M_j}$.
}
\begin{eqnarray}
V_{ij}^{(\text{TW})}(\sqrt{s}) = 
- {C^{(\text{TW})}_{ij}\over 8 f_{i}f_{j}}\,{\cal N}_i {\cal N}_j (2\sqrt{s} - M_i - M_j)~~,
\label{eq:TW}
\end{eqnarray}
where $f_{i}$, $M_i$ and $E_i = \sqrt{M_i^2 + q_i^2}$ are the meson decay constant, the (physical) baryon mass and the baryon energy in channel $i$. The center-of-mass three-momentum $q_i$ in that channel is: 
\begin{eqnarray}
q_i =
\frac{ \sqrt{ [ s-(M_i+m_i)^2 ] [ s-(M_i-m_i)^2 ] } }{2\sqrt{s}}~~,
\end{eqnarray}
with the meson mass $m_i$. The normalization factor is ${\cal N}_i = \sqrt{M_i+E_i}$. The constants $C^{(\text{TW})}_{ij}$ are determined by SU(3) Clebsch-Gordan coefficients and given in  Refs.~\cite{Kaiser1995, Borasoy2005}. 

Next in the ChPT hierarchy are the Born terms derived from the meson-baryon Yukawa vertices in Eq. (\ref{eq1}). The direct Born term shown in Fig.~\ref{Fig1}(ii) is given by
\begin{eqnarray}
V^{(D)}_{ij}(\sqrt{s}) =
-\sum_{k=1}^{8}
\frac{C^{(\text{Born})}_{\bar{i}i,k}  C^{(\text{Born})}_{\bar{j}j,k}}{12 f_{i}f_{j}}
\,{\cal N}_i\,{\cal N}_j\,
\frac{(\sqrt{s}-M_i)(\sqrt{s}-M_k)(\sqrt{s}-M_j)}{s-M_k^2}~~,
\label{eq:directBorn}
\end{eqnarray}
with channel $k$ denoting an intermediate baryon. The coefficients $C^{(\text{Born})}_{\bar{i}i,k}$ include the constants $D$ and $F$ and are tabulated in Ref.~\cite{Borasoy2005}. The crossed Born term (Fig.~\ref{Fig1}(iii)) 
involves the u-channel Mandelstam variable $u$  and is given by
\begin{align}
V^{(C)}_{ij}(\sqrt{s}) =&
\sum_{k=1}^{8}
\frac{C^{(\text{Born})}_{\bar{j}k,i} C^{(\text{Born})}_{\bar{i}k,j}}{12 f_{i}f_{j}}\,
{\cal N}_i \,{\cal N}_j\,
\frac{1}{u - M_k^2}  \nonumber\\
& \times
\biggl[
    \sqrt{s}+M_k
    -\frac{(M_i+M_k)(M_j+M_k)}{2N_i^2N_j^2}
    (\sqrt{s}-M_k+M_i+M_j) \nonumber\\*
    &+\frac{(M_i+M_k)(M_j+M_k)}{4q_iq_j}
    \bigl\{
    \sqrt{s}+M_k-M_i-M_j \nonumber\\*
    &-\frac{s+M_k^2-m_i^2-m_j^2-2E_iE_j}{2N_i^2N_j^2}
    (\sqrt{s}-M_k+M_i+M_j)\bigr\} \nonumber \\
    &\times
    \ln\frac{s+M_k^2-m_i^2-m_j^2-2E_iE_j-2q_iq_j}
    {s+M_k^2-m_i^2-m_j^2-2E_iE_j+2q_iq_j}
    \biggr]~~.
\label{eq:crossedBorn}
\end{align}

Finally we turn to the next-to-leading oder terms (Fig.~\ref{Fig1}(iv)). The corresponding interactions are derived from the four-point vertices in Eq.~(\ref{eq2}): 
\begin{eqnarray}
V^{(\text{NLO})}_{ij}(\sqrt{s}) =
\frac{{\cal N}_i\, {\cal N}_j}{f_{i}f_{j}}\, 
\Bigl[
C_{ij}^{(\text{NLO1})} - 2 C_{ij}^{(\text{NLO2})}
\left(
E_i E_j + \frac{q_i^2 q_j^2}{3 {\cal N}_i\,{\cal N}_j}
\right)
\Bigr]~~.
\label{eq:NLO}
\end{eqnarray}
The coefficients $C^{(\text{NLO1})}_{ij}$ and $C^{(\text{NLO2})}_{ij}$ are again summarized in Ref.~\cite{Borasoy2005}. They include the NLO low-energy constants $b_0$, $b_D$, $b_F$ and $d_i$. Since we take into account the renormalized (physical) masses of the baryons, these low-energy constants are not identical to the ones from tree-level ChPT.

\subsection{Chiral coupled-channels dynamics}

In contrast to the SU(2) pion-nucleon systems close to threshold, the meson-baryon channels in the strangeness sector are strongly interacting. In particular, the $\bar{K}N$ interaction is sufficiently strong to produce the $\Lambda(1405)$ as a quasi-bound state below the $\bar{K}N$ threshold. In such a situation, a perturbative calculation does not work and a nonperturbative resummation is mandatory in order to account for the strong coupled-channels dynamics.  The meson-baryon interactions derived above are thus used as the interaction matrix kernel ${\bf V}$ of the coupled-channels Bethe-Salpeter equation for the $\mathbf{T}$-matrix: 
\begin{eqnarray}
\mathbf{T} = \mathbf{V} + \mathbf{V\cdot G \cdot T} = (\mathbf{V}^{-1} - \mathbf{G})^{-1}~~.
\label{tmatrix}
\end{eqnarray}
Here $\bf{G}$ is a diagonal matrix, $G_{ij} = G_i(Q)\,\delta_{ij}$, with elements $G_i$ representing the meson-baryon loop function in channel $i$: 
\begin{eqnarray}
G_i(Q) = \int {d^4k \over (2\pi)^4}{\textrm{i}\over [(Q-k)^2 - M_i^2 + \textrm{i}\epsilon]
(k^2-m_i^2  + \textrm{i}\epsilon)} ~~.
\end{eqnarray}
Its logarithmic divergence can be tamed by dimensional regularization as
\begin{align}
G_i(\sqrt{s}) = &
a_i(\mu) +  {1\over 32\pi^2}\left[\ln\left({m_i^2\,M_i^2\over \mu^4}\right) - {M_i^2 - m_i^2\over s}\ln\left({m_i^2\over M_i^2}\right)\right] \nonumber\\
&- {1\over 16\pi^2}\left[1 + {4q_i\over\sqrt{s}}\,\textrm{artanh}\left({2\sqrt{s}\,q_i \over (m_i+M_i)^2 -s}\right)\right]~~.
\label{eq:G}
\end{align}
The subtraction constants $a_i(\mu)$ act as renormalization parameters at a scale $\mu$ such that the $\mathbf{T}$-matrix (\ref{tmatrix}) is scale independent. 
Note that the last expression in Eq. (\ref{tmatrix}) is well justified in formal scattering theory, e.g. by using the N/D method which automatically guarantees the unitarity of the scattering amplitude. As demonstrated in Refs. \cite{Oller2001,HJ2011} the interaction kernel ${\bf V}$ can be identified with the one  
derived from tree-level ChPT up to next-to-leading order with no problem of double counting.

\subsection{Observables} 

The meson-baryon coupled-channels $\mathbf{T}$-matrix is the starting point for the calculation of various $K^{-}p$ scattering observables. This $\mathbf{T}$-matrix is related to the forward scattering amplitudes $f_{ij}$ as
\begin{eqnarray}
f_{ij}(\sqrt{s}) = {1\over 8\pi\sqrt{s}}\,T_{ij}(\sqrt{s})~~.
\end{eqnarray}
The $K^-p$ elastic scattering amplitude at threshold defines the scattering length, $a(K^-p) = f_{11}(\sqrt{s} = m_{K^-}+M_p)$, a complex number because of the absorptive channels converting $K^-p$ into $\pi\Sigma$ and $\pi\Lambda$. The energy shift and width of the 1s state of kaonic hydrogen are related to the $K^-p$ scattering length, with important second order corrections, as follows~\cite{MRR2004}:
\begin{eqnarray}
\Delta E - \textrm{i}\Gamma/2 = -2\alpha^3\,\mu_r^2\,a(K^-p)\left[1+2\alpha\,\mu_r\,(1-\ln\alpha)\,a(K^-p)\right]~~,
\label{eq10}
\end{eqnarray}
where $\alpha$ is the fine-structure constant and the $K^-p$ reduced mass is given by $\mu_r = m_{K^{-}} M_p/(m_{K^{-}} + M_p)$. 

The total reaction cross sections in the various meson-baryon scattering channels are given by
\begin{eqnarray}
\sigma_{ij}(\sqrt{s}) = {q_i\over q_j}{|T_{ij}(\sqrt{s})|^2\over 16\pi\,s}~~,
\end{eqnarray}
where the cross section is defined for  $\sqrt{s}> M_{i}+m_{i}$, above the threshold of the final-state channel $i$. For the $K^-p$ elastic cross section, we also take into account electromagnetic interactions which are important near the $K^- p$ threshold~\cite{Borasoy2005}. The Coulomb interaction gives an additional contribution to the diagonal amplitude in the $K^{-}p$ channel:
\begin{eqnarray}
f^{\rm{Coul}}_{11}(\sqrt{s},\theta_{\rm cm})& = &
\frac{1}{2 q_1^2\,a_B\, \sin^2(\theta_{\rm cm}/2)} \nonumber \\
& &\times \,\,\frac{\Gamma(1-\textrm{i}/(q_1\,a_B))}{\Gamma(1+\textrm{i}/(q_1\, a_B))}
\,\,\exp\Bigl(\frac{2 \textrm{i}}{q_1\,a_B} \ln \sin\frac{\theta_{\rm cm} }{2}\Bigr)~~,
\end{eqnarray}
with $a_B = 84$ fm, the Bohr radius of the $K^- p$ system, and $\theta_{\rm cm}$ denoting scattering angle. This Coulomb amplitude is added to the strong interaction amplitude and the scattering angle is integrated up to $\cos \theta_{\rm cm} < 0.966$ to avoid the divergence at $\theta_{\rm cm} = 0$. 

Several combinations of $K^{-}p$ inelastic yields at threshold are known in the form of branching ratios defined as
\begin{align}
\gamma =& {\Gamma(K^-p\rightarrow\pi^+\Sigma^-)\over\Gamma(K^-p\rightarrow\pi^-\Sigma^+)} = {\sigma_{51}\over\sigma_{61}}~,~~R_n = {\Gamma(K^-p\rightarrow\pi^0\Lambda)\over\Gamma(K^-p\rightarrow\textrm{neutral states})} 
= {\sigma_{31}\over\sigma_{31} + \sigma_{41}}~,\nonumber\\
R_c =& {\Gamma(K^-p\rightarrow\pi^+\Sigma^-,\,\pi^-\Sigma^+)\over\Gamma(K^-p\rightarrow\textrm{all inelastic channels})}= {\sigma_{51}+\sigma_{61}\over \sigma_{31}+\sigma_{41}+\sigma_{51}+\sigma_{61}}~~,
\label{branchings}
\end{align}
with all partial cross sections $\sigma_{ij}$ calculated at the $K^-p$ threshold.

\begin{table}[htbp]
  \begin{center}
   \begin{tabular}{l|l|l|l|l}  
      $\Delta E$ [eV]  & $\Gamma$ [eV] & $\gamma$ & $R_n$ & $R_c$ \\ \hline
 $283\pm 36 \pm 6$   & $541\pm 89 \pm 22$ & 
 $2.36\pm 0.04$ & $0.189\pm 0.015$ & $0.664\pm 0.011$
    \end{tabular}
    \caption{
    Experimental observations of the energy shift and width of the 1s state of kaonic hydrogen ($\Delta E$ and $\Gamma$)~\cite{Bazzi2011}, 
    threshold branching ratios ($\gamma$, $R_{n}$ and $R_{c}$)~\cite{BR}.
    }
    \label{tab:experiment}
  \end{center}
\end{table}
\section{Results and discussion}
\subsection{Fitting procedure}

We now describe the systematic fitting procedure used in the framework of the chiral SU(3) dynamics at NLO level. We first summarize the empirical constraints that enter this study. Important constraints are the kaonic hydrogen shift and width from the SIDDHARTA measurements~\cite{Bazzi2011}: 
\begin{eqnarray}
\Delta E = 283\pm36(stat)\pm6(syst)~ \textrm{eV} ~, ~~~\Gamma = 541\pm89(stat)\pm22(syst)~\textrm{eV}~~.\nonumber 
\end{eqnarray}
The threshold branching ratios (\ref{branchings}) are determined~\cite{BR} from $K^{-}$ capture on hydrogen as listed in Table~\ref{tab:experiment}. In addition to these $K^{-}p$ threshold constraints, we also make use of the total cross section data accumulated in Refs.~\cite{cross_sections} (see Fig.~\ref{Fig2}). In contrast to several previous studies, 
the currently available $\pi\Sigma$ mass spectra are not included in the empirical data base. At this point a meaningful comparison with experimental data would require a detailed investigation of the different reaction mechanisms generating such spectra in each given experiment~\cite{HJ2011}. Instead, the $\pi\Sigma$ spectrum emerges here as a prediction following the fitting to the previously mentioned quantities.

The $\chi^2$ fits to the data base have been performed using three consecutive schemes, systematically improving the interaction kernel in each step. The first setup involves just the Tomozawa-Weinberg (TW) term~\eqref{eq:TW}, the dominant component of the leading-order interaction. In the second step we include in additon the direct and crossed Born terms, Eqs.~\eqref{eq:directBorn} and \eqref{eq:crossedBorn}, completing the $\mathcal{O}(p)$ ChPT interaction (TWB). The third step incorporates all terms~\eqref{eq:NLO} of the full NLO model. We define the overall $\chi^2$ following Ref.~\cite{Borasoy2006}. Consider a measurement, labeled $i$, and the number of data points $n_i$ associated with this measurement. We first calculate $\chi^2_i$ for each $i$-th measurement, then collect all $\chi^2_i$ multiplied by proper weight factors, and finally obtain overall $\chi^2$/d.o.f.,
\begin{eqnarray}
\chi^2 \textrm{/d.o.f.} = 
\frac{\sum_{i=1}^{R} n_i}{R (\sum_{i=1}^{R} n_i - p ) }\sum_{i=1}^{R} \frac{\chi^2_i}{n_i},
\label{chi2}
\end{eqnarray}
where $R$ is the total number of measurements considered. Furthermore, $p$ is the number of parameters that appears in each step, increasing from TW via TWB to the full NLO setup. If all measurements have same numbers of data points, Eq.~(\ref{chi2}) reduces to the standard definition of $\chi^2$/d.o.f.

In the numerical calculations we use the physical masses for mesons and baryons. This is necessary in order to reproduce the correct threshold energies in the different meson-baryon channels. The constants associated with the axial vector baryon couplings are given by $D=0.80$ and $F=0.46$ (i.e. $g_A = D+F = 1.26$). The meson decay constants are chosen at their physical values~\cite{decay_constants}
\begin{eqnarray}
f_\pi = 92.4~\textrm{MeV}~, ~~~f_K = (1.19\pm0.01)\,f_\pi~,~~~f_\eta = (1.30\pm0.05)\,f_\pi~.
\label{eq:decayconst}
\end{eqnarray}
Using the physical masses and decay constants is justified by taking into account the renormalization of the baryon masses and meson fields. As a consequence, parts of the effects of the NLO parameters $b_0$, $b_D$ and $b_F$ have already been absorbed in the renormalized quantities, shifting the baryon octet masses from their degenerate chiral limit, $M_0$, to their physical values; the parameters used in the interaction kernel are therefore to be interpreted as the renormalized ones. We denote these renormalized NLO parameters as $\bar{b}_0$, $\bar{b}_D$ and $\bar{b}_F$.  They are expected to be considerably smaller in magnitude than the ones usually quoted in tree-level chiral perturbation theory. 

Thus, the free parameters to be determined by the $\chi^2$ fits are:\\
 {\bf i}) the subtraction constants $a_i(\mu)$ that are assumed to be isospin symmetric; \\
 {\bf ii}) the low energy constants in the NLO Lagrangian $\bar b_0$, $\bar b_D$, $\bar b_F$ and $d_i$.\\
 Note that the set i) appears in all three TW, TWB and NLO versions,  whereas set ii) is only used in the full NLO scheme. In the fitting procedure we also allow for small modifications of the $K$ and $\eta$ decay constants  within the uncertainties given in Eq.~\eqref{eq:decayconst}.

\begin{table}[htbp]
  \begin{center}
    \begin{tabular}{l|lll}  
               & ~~~~~TW & ~~~~TWB & ~~~~~NLO \\
      \hline \\
      ~~$a_{\bar{K}N}$    ($10^{-3}$) &~~~$-1.57$  &~~~$-1.04$ &~~~\,\,$-2.38$  \\
      ~~$a_{\pi\Lambda}$  ($10^{-3}$) &$-107.97$&~~~$-8.06$ &~~\,$-16.57$ \\
      ~~$a_{\pi\Sigma}$   ($10^{-3}$) &~~~~~\,$2.31$   &~~~~~\,$2.96$ &~~~~~~\,$4.35$   \\
      ~~$a_{\eta\Lambda}$ ($10^{-3}$) &~~~$-0.20$  &~~~$-3.46$ &~~~~$-0.01$  \\
      ~~$a_{\eta\Sigma}$  ($10^{-3}$) &~~\,$216.37$ &~~~~~\,$3.52$  &~~~~~~\,$1.90$   \\
      ~~$a_{K \Xi}$       ($10^{-3}$) &~~~~$39.48$  &~~~~\,$12.51$ &~~~~~$15.83$  \\ \\
      \hline \\
      ~~~$f_K$ (MeV)                  &~~~~$110.8$ &~~~~$109.0$ &~~~~~$110.0$ \\
      ~~~$f_{\eta}$ (MeV)             &~~~~$124.5$ &~~~~$124.6$ &~~~~~$118.8$ \\ \\
      \hline \\
      ~~$\bar{b}_{0}$  ($10^{-2}$ GeV$^{-1}$) &~~~~~~$-$ &~~~~~~$-$ &~~~\,$-4.79$  \\
      ~~$\bar{b}_{D}$  ($10^{-2}$ GeV$^{-1}$) &~~~~~~$-$ &~~~~~~$-$ &~~~~~~$0.48$ \\
      ~~$\bar{b}_{F}$  ($10^{-2}$ GeV$^{-1}$) &~~~~~~$-$ &~~~~~~$-$ &~~~~~~$4.01$   \\
      ~~$d_{1}$  ($10^{-2}$ GeV$^{-1}$)       &~~~~~~$-$ &~~~~~~$-$ &~~~~~~$8.65$  \\
      ~~$d_{2}$  ($10^{-2}$ GeV$^{-1}$)       &~~~~~~$-$ &~~~~~~$-$ &~~$-10.62$   \\
      ~~$d_{3}$  ($10^{-2}$ GeV$^{-1}$)       &~~~~~~$-$ &~~~~~~$-$ &~~~~~~$9.22$  \\
      ~~$d_{4}$  ($10^{-2}$ GeV$^{-1}$)       &~~~~~~$-$ &~~~~~~$-$ &~~~~~~$6.40$  \\ \\
      \hline \\
      ~~$\chi^{2}$/d.o.f. & ~~~~~$1.12$ & ~~~~~$1.15$ & ~~~~~~$0.96$   \\
    \end{tabular}
    \caption{
   Parameters resulting from the systematic $\chi^{2}$ analysis, 
    using leading order (TW) plus Born terms (TWB) and full NLO schemes.
    Shown are the isospin symmetric subtraction constants $a_i(\mu)$ at $\mu=1$ GeV,
    the meson decay constants $f_K$ and $f_{\eta}$,
    the renormalized NLO constants $\bar{b}_i$ and $d_i$, 
    and $\chi^{2}$/d.o.f. of the fit.}
    \label{tab:parameters}
  \end{center}
\end{table}
\begin{table}[htbp]
  \begin{center}
   \begin{tabular}{l|lll}  
               & ~~~~~~TW & ~~~~~TWB & ~~~~NLO \\ \\
      \hline \\
      $\Delta E$ [eV]  &~~~~~~373 & ~~~~~~\,377 & ~~~~~\,306 \\
      $\Gamma$ [eV]    &~~~~~~495 & ~~~~~~\,514 & ~~~~~\,591 \\ \\
      $\gamma$    &~~~~~2.36 & ~~~~~~2.36 & ~~~~~2.37  \\
      $R_{n}$     &~~~~ 0.20 & ~~~~~~0.19 & ~~~~~0.19  \\
      $R_{c}$    &~~~~ 0.66 & ~~~~~~0.66 &~~~~~0.66    \\  \\
      \hlineÊ\\
      pole positions  ~~& ~$1422-16$\,i ~~& ~~$1421-~17$\,i ~~&~~$1424-26$\,i~~   \\
      ~~~~~$[\textrm{MeV}]$  &~$1384-90$\,i~~ &~~$1385-105$\,i~~ &~~$1381-81$\,i~~  \\
    \end{tabular}
    \caption{
    Results of the systematic $\chi^{2}$ analysis using leading order (TW) 
    plus Born terms (TWB) and full NLO schemes. 
    Shown are 
    the energy shift and width of the 1s state of kaonic hydrogen ($\Delta E$ and $\Gamma$), 
    threshold branching ratios ($\gamma$, $R_{n}$ and $R_{c}$), 
    and the pole positions of the isospin $I=0$ amplitude in the $\bar{K}N$-$\pi\Sigma$ domain.}
    \label{tab:results}
  \end{center}
\end{table}

\subsection{Fit results}

\begin{figure}
\includegraphics*[width=7cm]{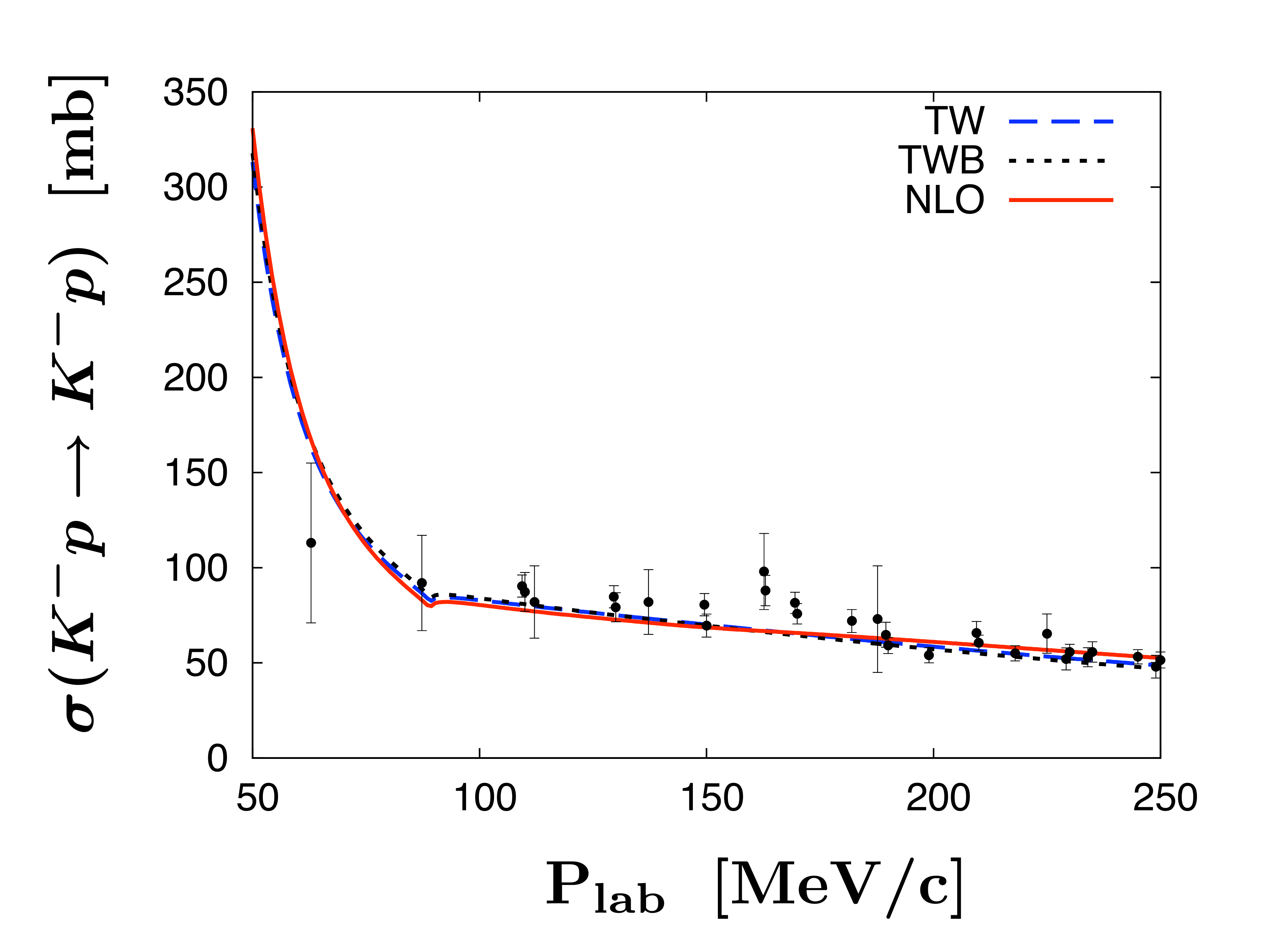}
\includegraphics*[width=7cm]{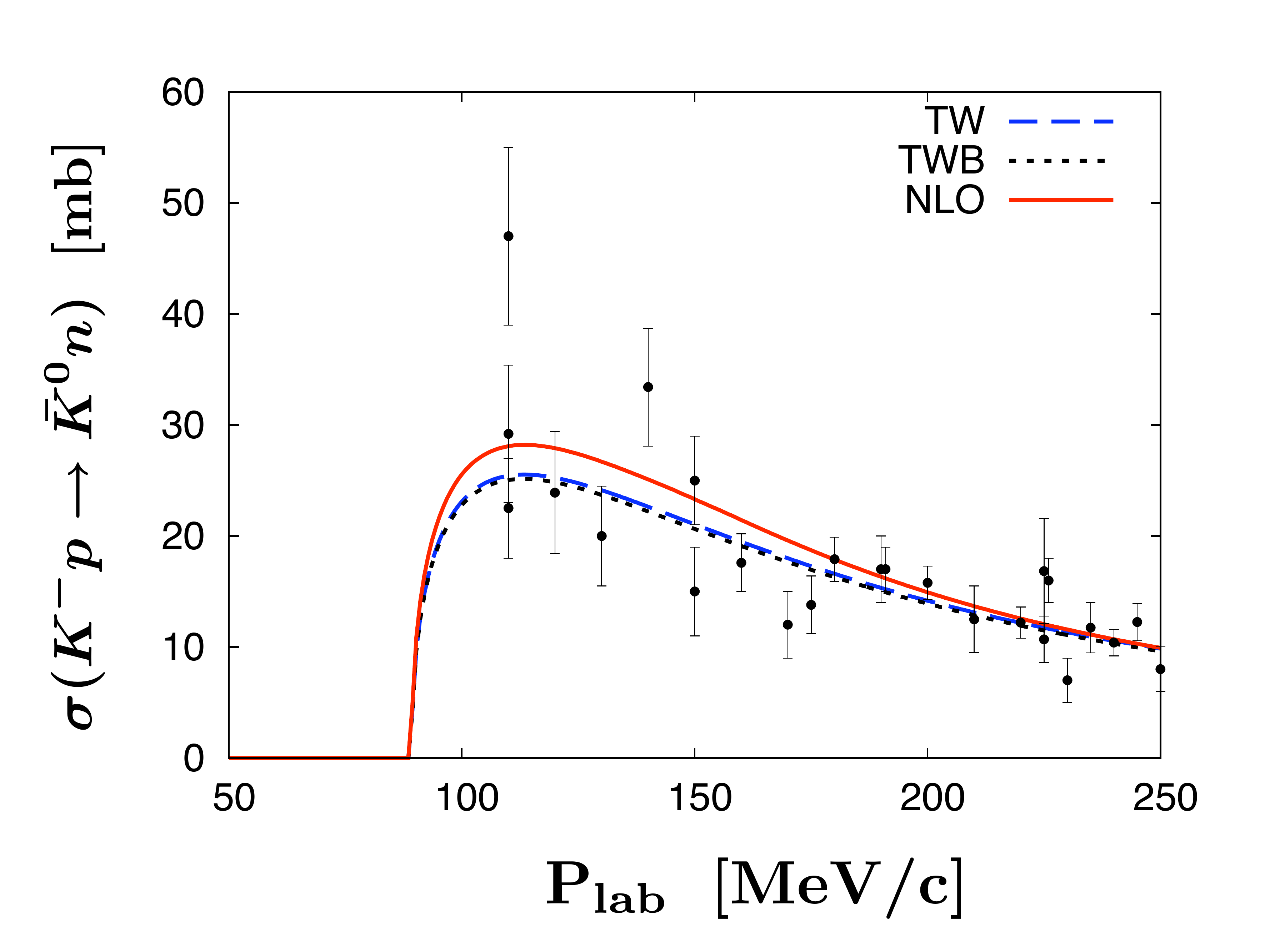}\\
\includegraphics*[width=7cm]{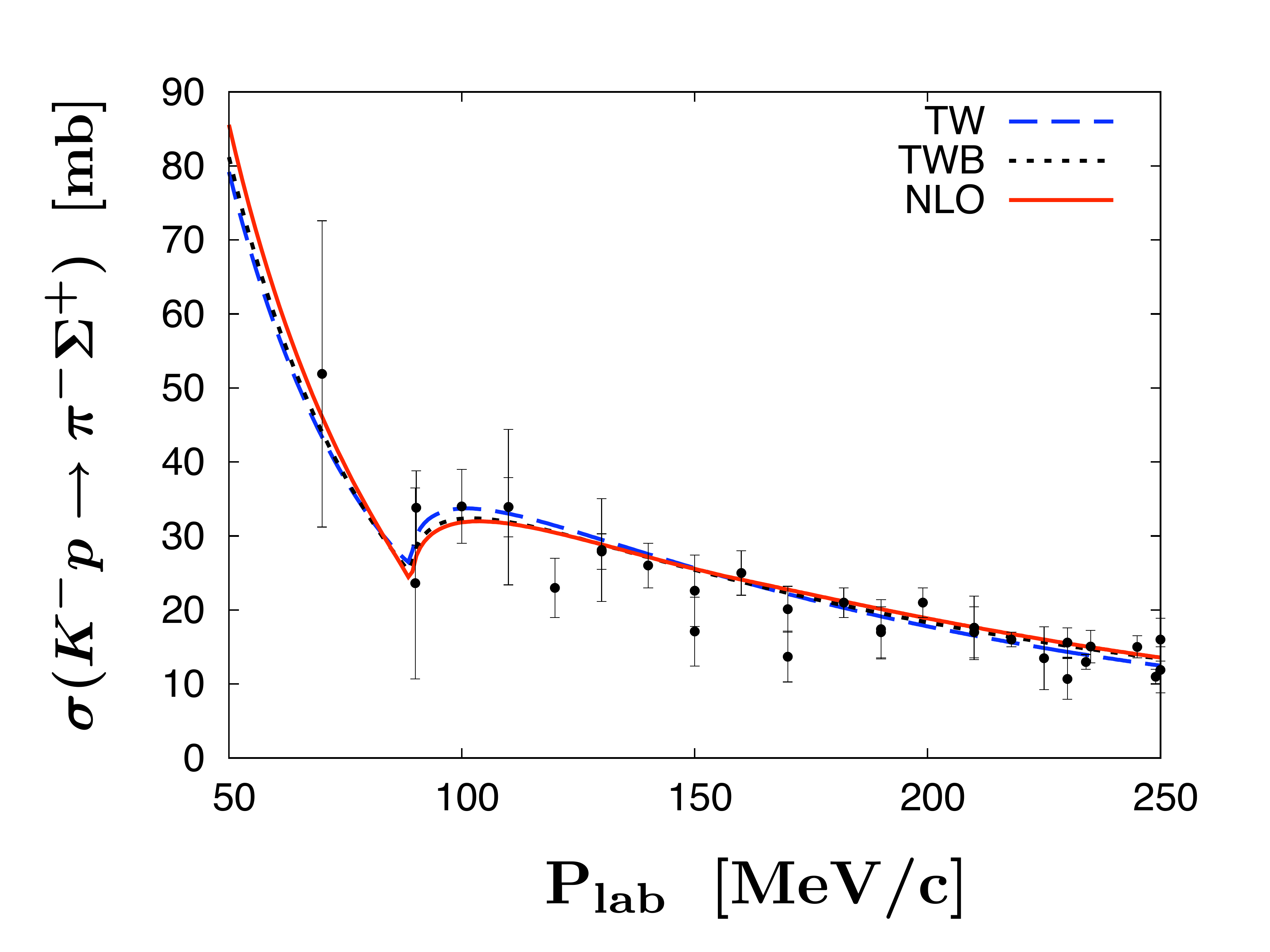}
\includegraphics*[width=7cm]{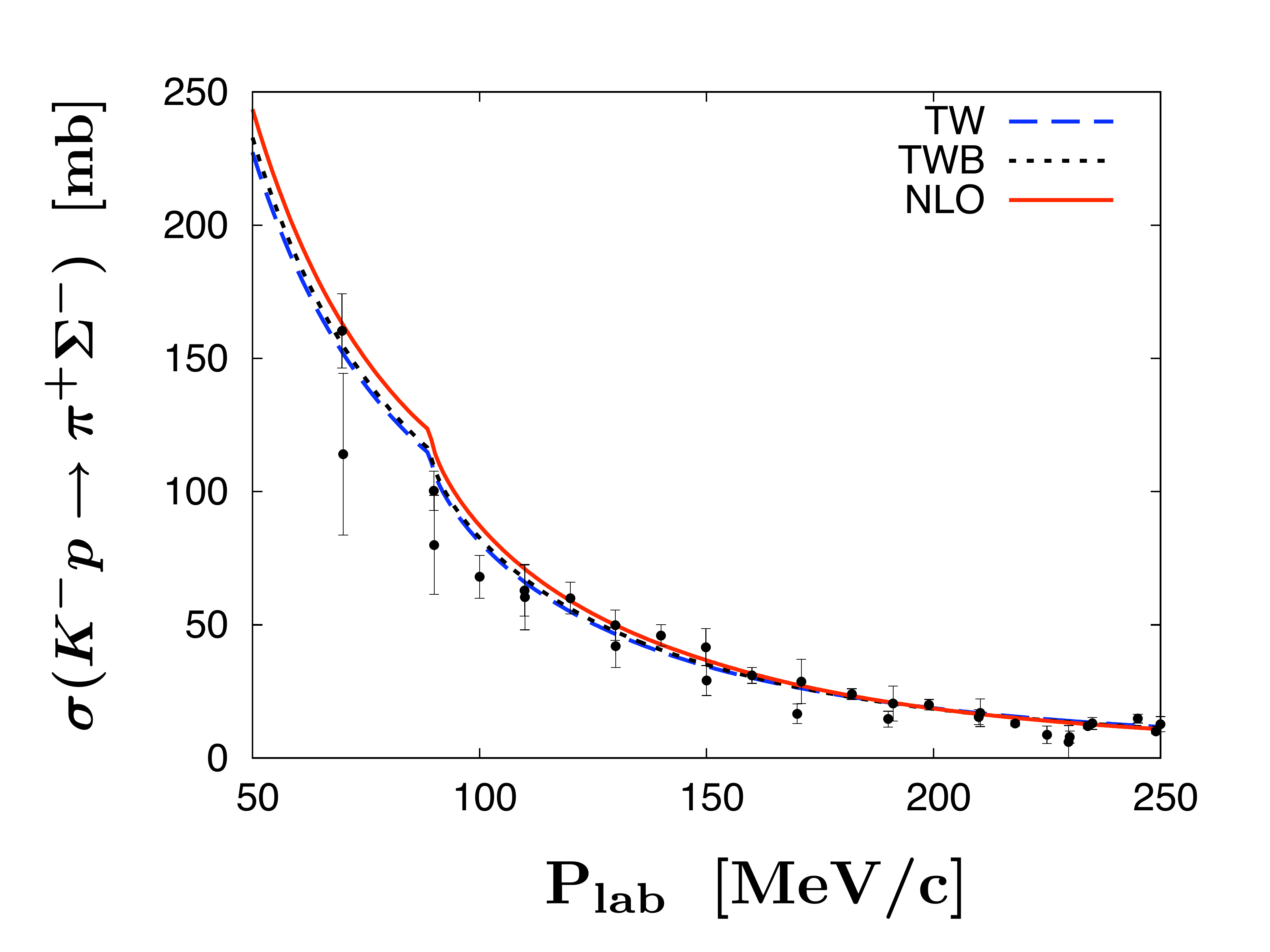}\\
\includegraphics*[width=7cm]{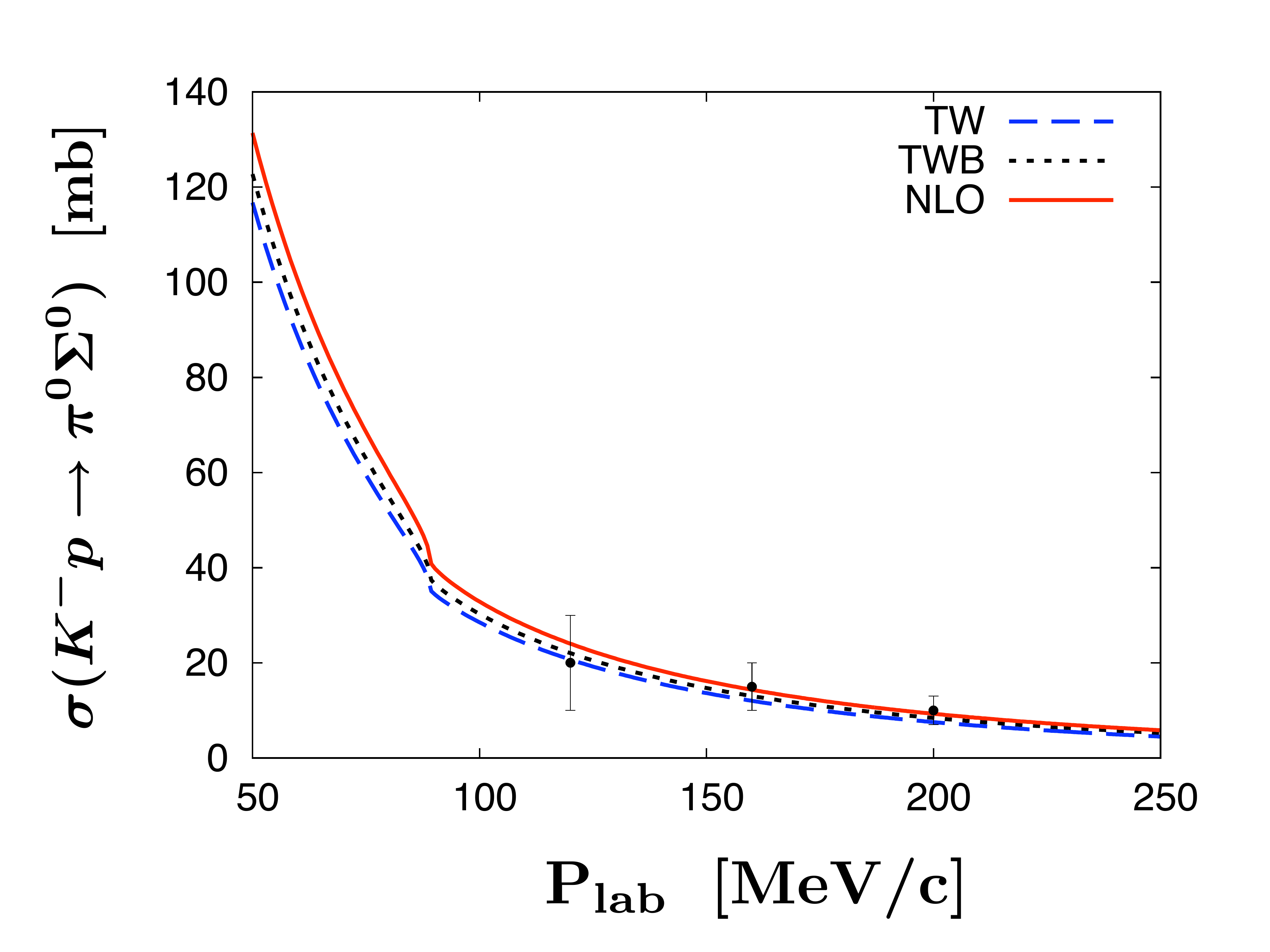}
\includegraphics*[width=7cm]{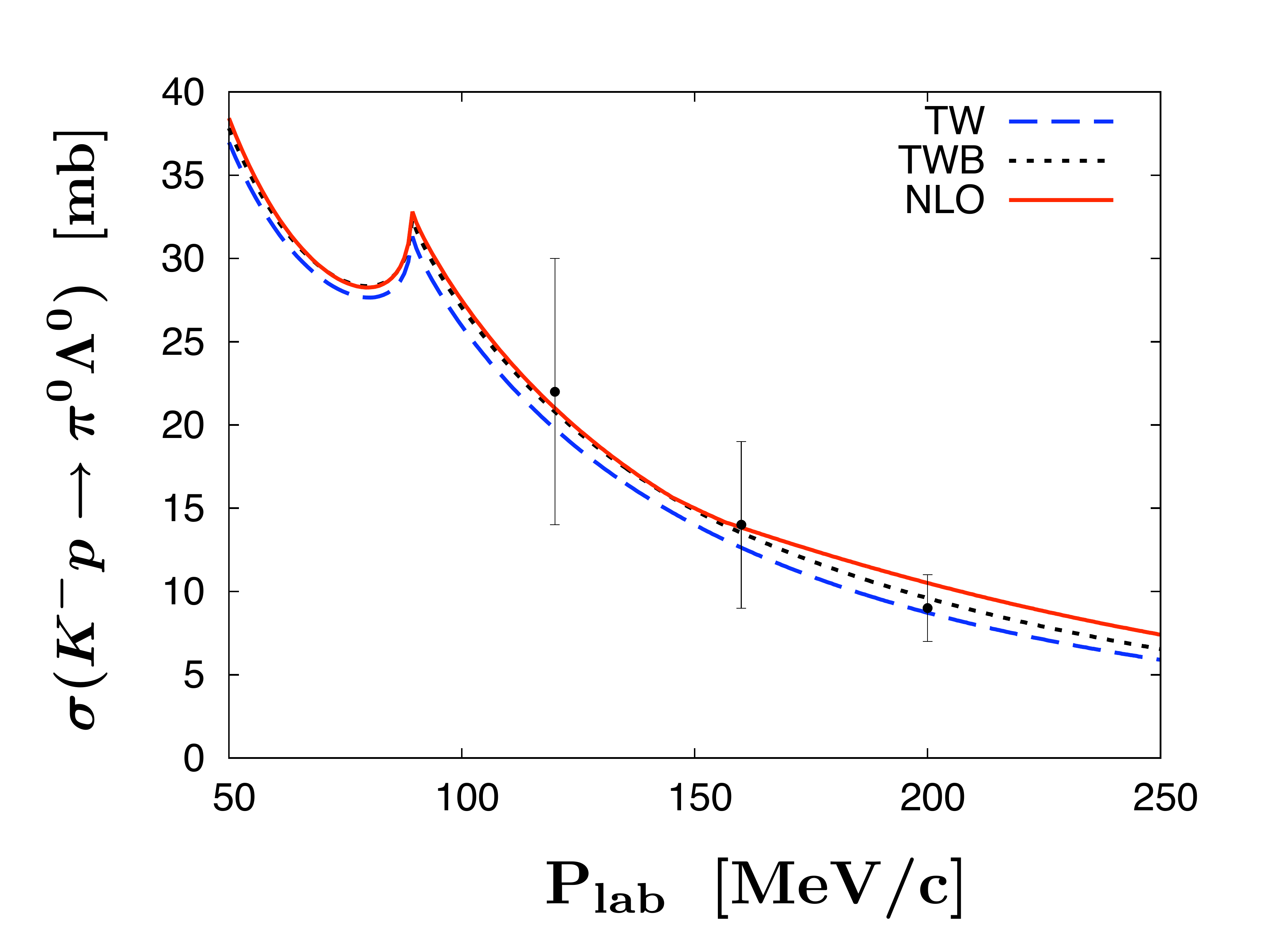}\\
\caption{Calculated $K^-p$ elastic, charge exchange and strangeness exchange cross sections as functions of $K^-$ laboratory momentum, compared with experimental data~\cite{cross_sections}. The dashed, dotted and solid curves represent TW, TWB and best fits of the full NLO calculations, respectively. 
}
\label{Fig2}
\end{figure}

With the TW terms alone a reasonable overall fit can already be reached with $\chi^2/\textrm{d.o.f.} = 1.12$, as shown in Table~\ref{tab:parameters}. However, although the branching ratios and cross sections are reproduced quite well as shown in Table~\ref{tab:results} and Fig.~\ref{Fig2}, the kaonic hydrogen energy shift ($\Delta E = 373$ eV) exceeds the empirical bound. It should also be noted that the fitted values of the subtraction constants become large in the $\pi\Lambda$ and $\eta\Sigma$ channels, beyond their expected ``natural" size, $|a|\sim 10^{-2}$~\cite{Oller2001,Hyodo:2008xr}. This result indicates the limit of applicability of the simple TW model for the quantitative discussion of the scattering amplitudes, given the remarkable accuracy that has been achieved in the SIDDHARTA measurements.

%
%

Next we consider the TWB scheme in which the direct and crossed Born terms are combined with the TW term. In this case, the $\chi^2/\textrm{d.o.f.}$ changes only marginally. The energy shift $\Delta E=377$ eV is not improved by the inclusion of the Born terms, but the values of the subtraction constants (Table~\ref{tab:parameters}) are now reaching ``natural" sizes. From the theoretical point of view, this now indicates a consistent description of the interaction kernel and the loop function.

Finally the NLO terms are added in the construction of the full amplitudes. The $\chi^{2}$ analysis provides the best fit parameters with $\chi^2/\textrm{d.o.f.} = 0.96$. The kaonic hydrogen shift and width are now within the error bars of the SIDDHARTA measurements. Threshold branching ratios and total cross sections of the $K^-p$ scattering and reaction processes are well reproduced as demonstrated in Table~\ref{tab:results} and Fig.~\ref{Fig2}. The parameters determined by the NLO fit are altogether meaningful: natural-sized subtraction constants and small renormalized NLO parameters. 
The stepwise improvement of the theoretical description of the meson-baryon amplitudes in the three schemes from TW via TWB to NLO is evident, emphasizing the important role of the accurate kaonic hydrogen data in constraining chiral SU(3) dynamics. In  contrast, the scattering data alone (see Fig.~\ref{Fig2}) do not provide a sensitive test for the different schemes.

\begin{figure}
\includegraphics*[width=7cm]{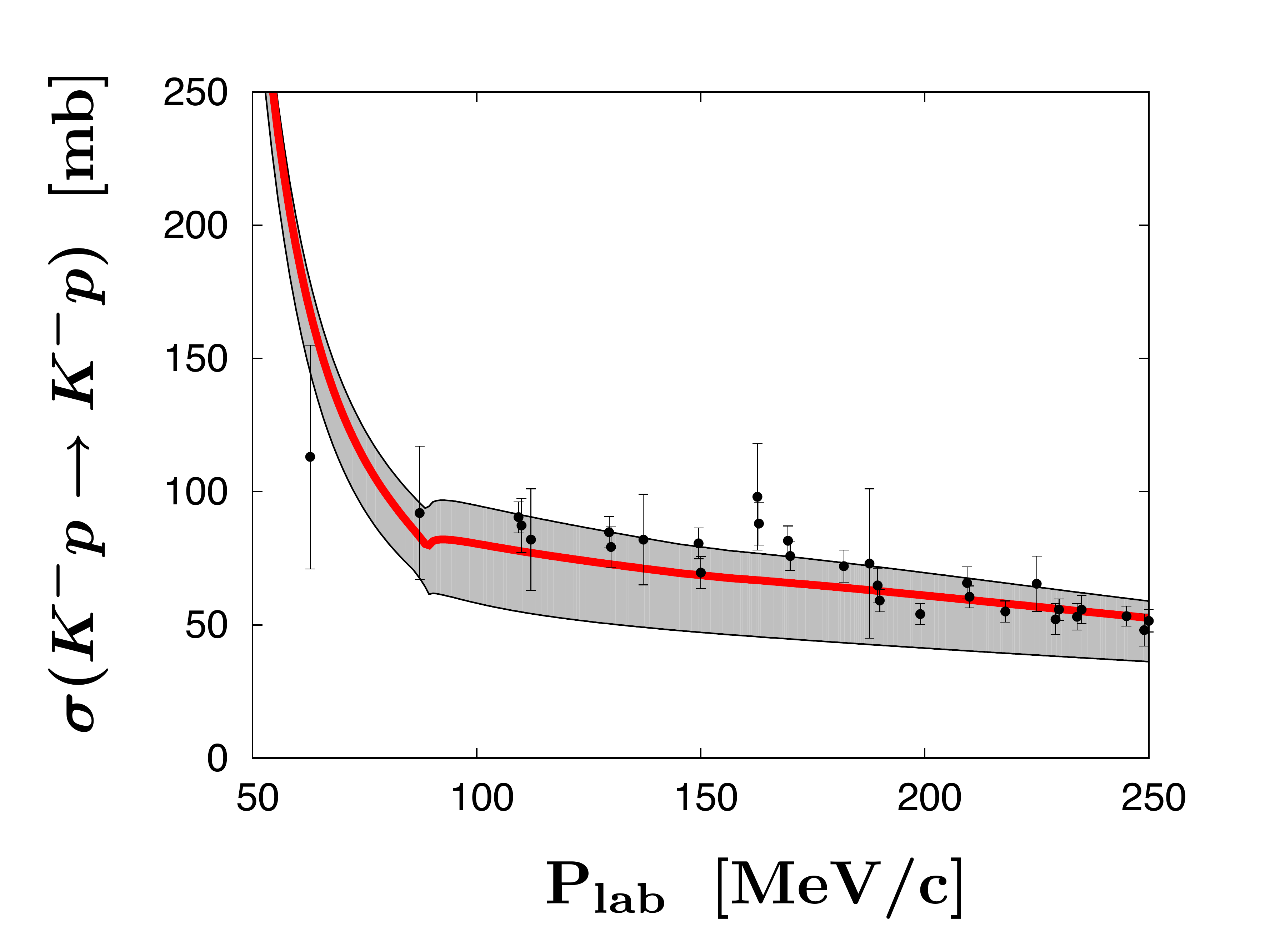}
\includegraphics*[width=7cm]{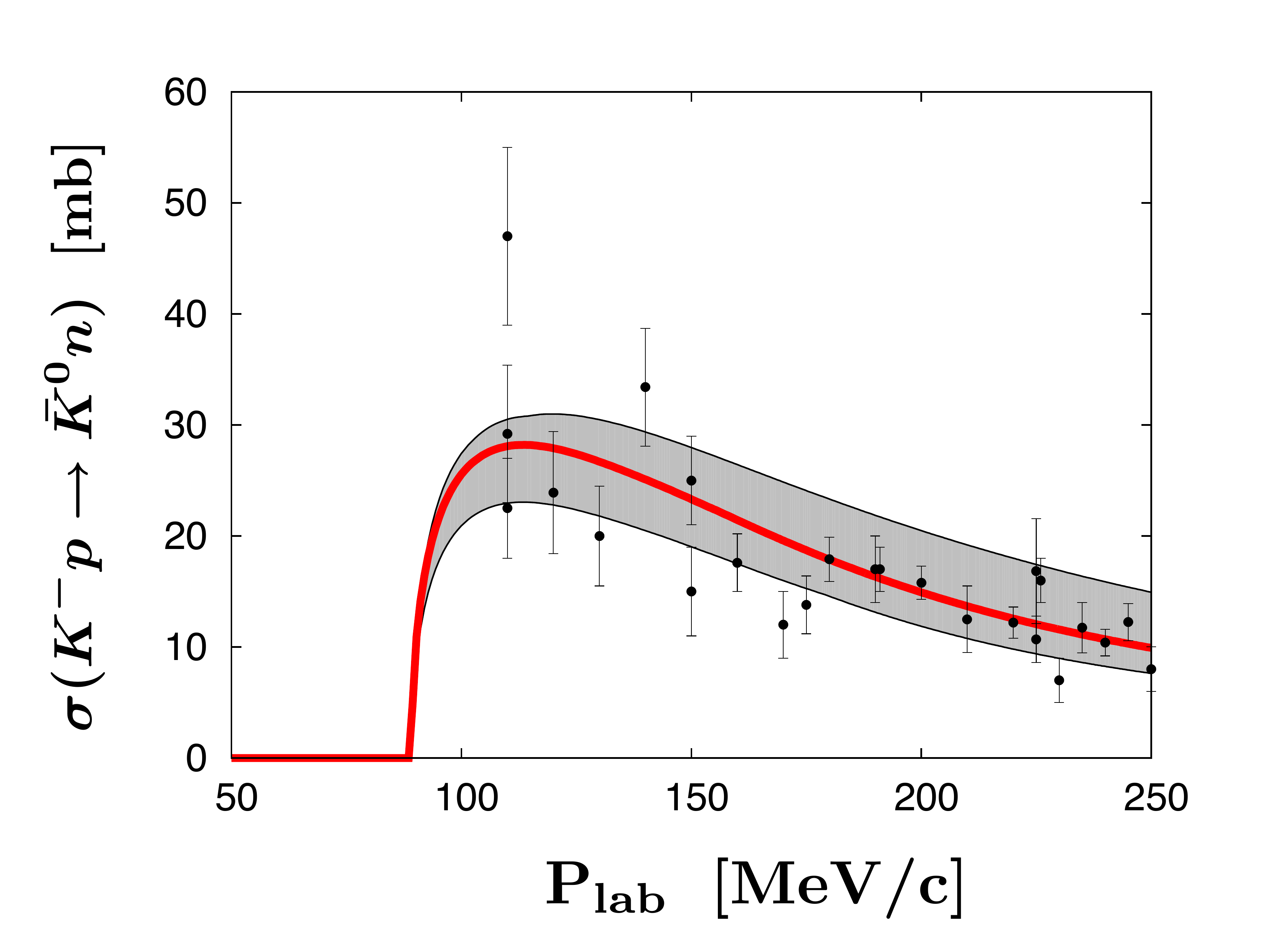}\\
\includegraphics*[width=7cm]{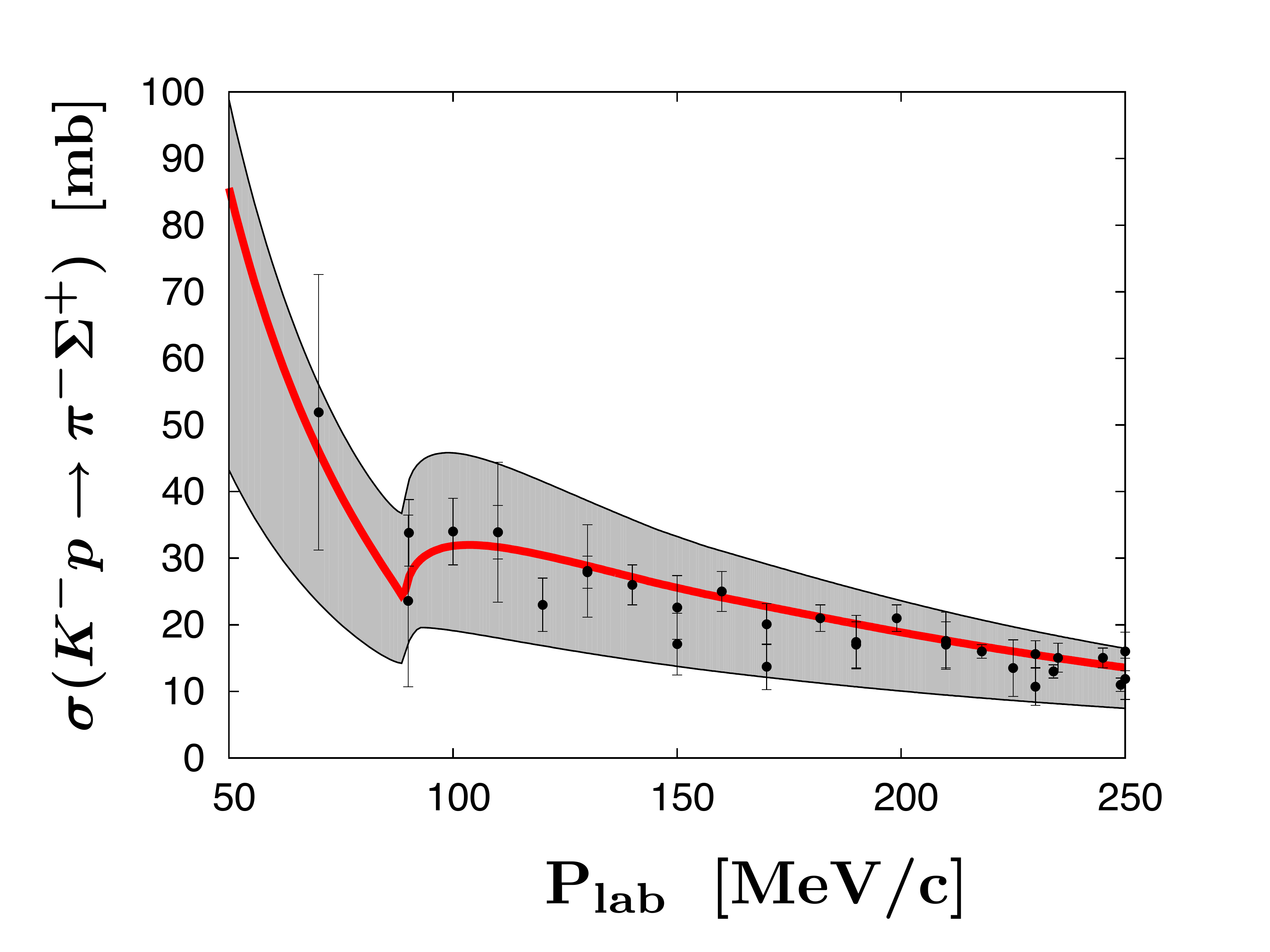}
\includegraphics*[width=7cm]{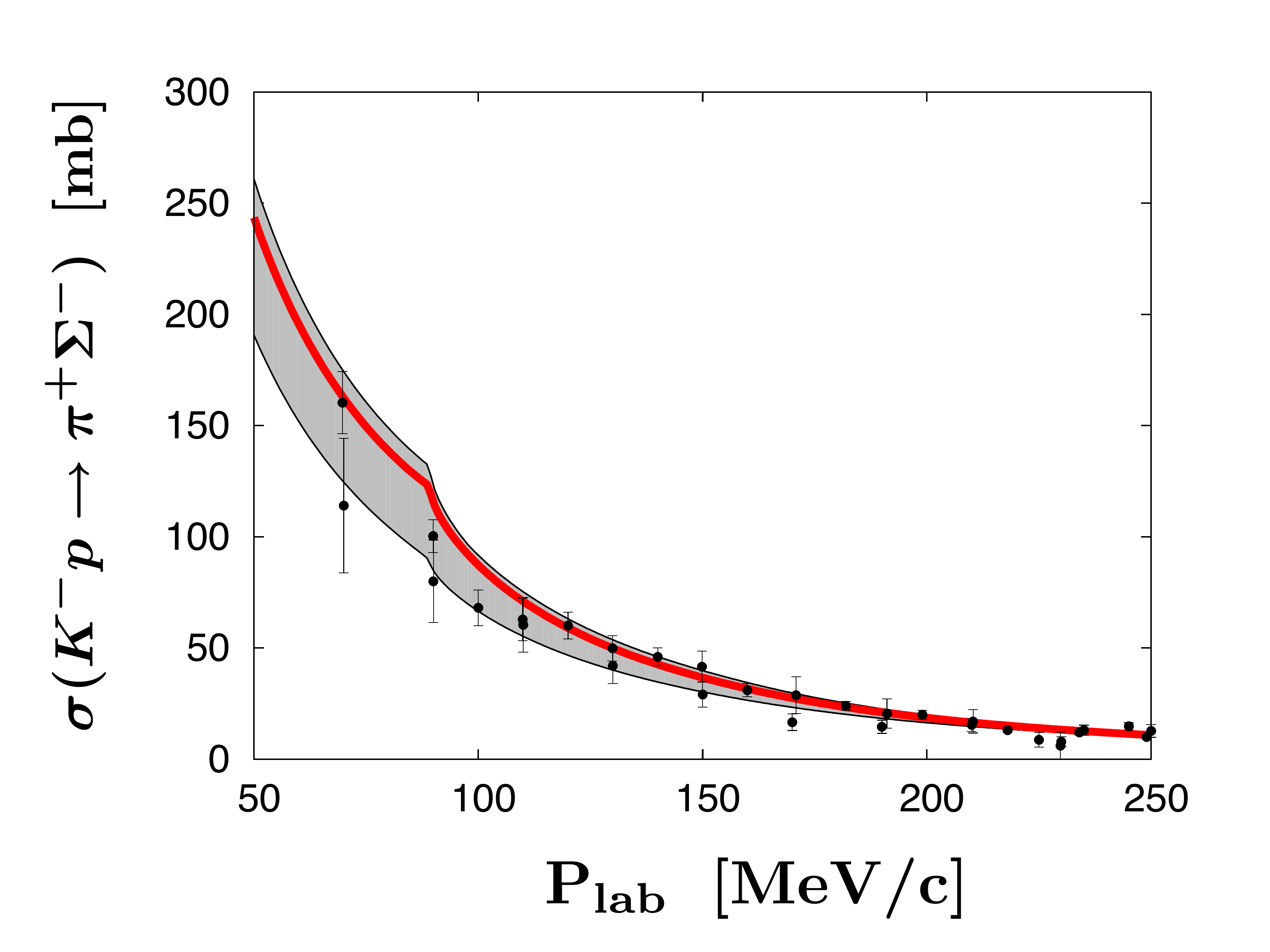}\\
\includegraphics*[width=7cm]{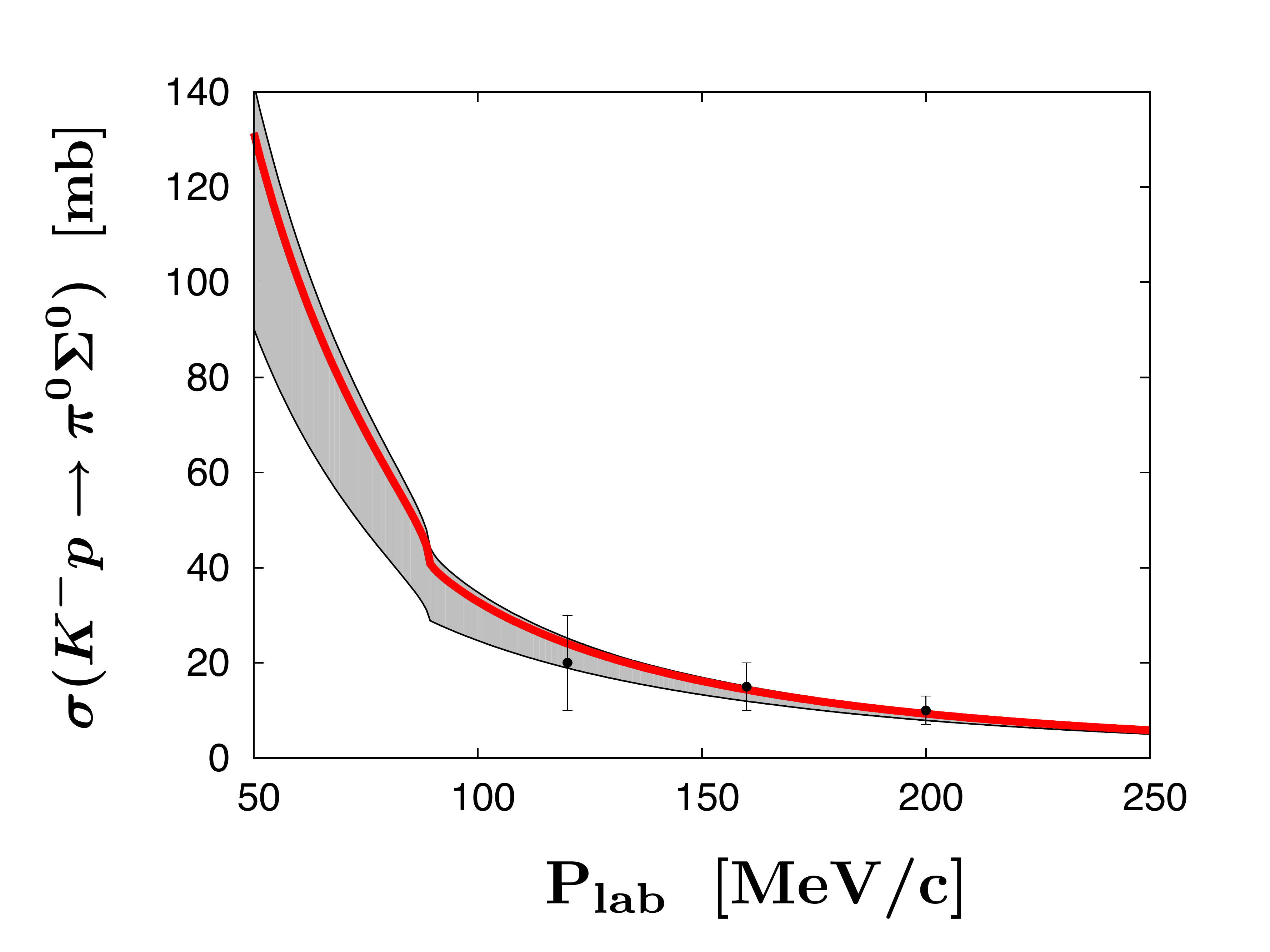}
\includegraphics*[width=7cm]{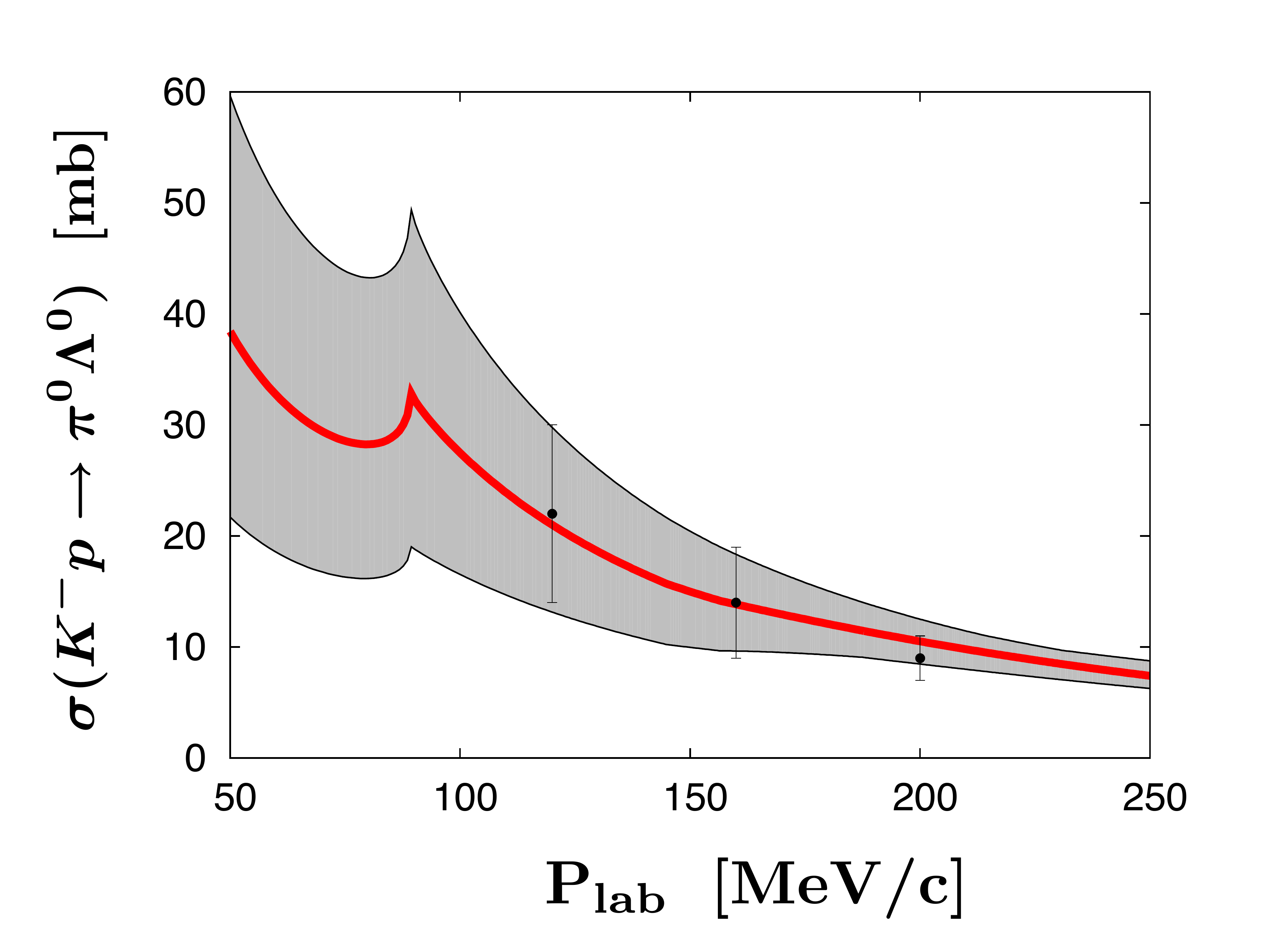}\\
\caption{Calculated $K^-p$ elastic, charge exchange and strangeness exchange cross sections
as functions of $K^-$ laboratory momentum, compared with experimental data~\cite{cross_sections}.  
The solid curves represent best fits of the full NLO calculations to the complete data base including threshold
observables. The shaded uncertainty bands are explained in the text.}
\label{Fig3}
\end{figure}

To estimate the uncertainty in the subthreshold extrapolation, we examine variations of the parameters around their best fit values. We vary the subtraction constants with the condition that the shift and width of the kaonic hydrogen are reproduced within experimental errors. In addition, the total cross section of the $K^-p \to \pi^0 \Lambda$ process is also used to set constraints on the $I=1$ amplitudes. With these error assignments, the allowed ranges for the subtraction constants are $a_i$ taken at the scale $\mu = 1$ GeV are found to be:
\begin{align}
    a_{\bar{K}N}=&-2.38^{+0.17}_{-0.84}~~, &
    a_{\pi \Lambda}=&-16.57^{+16.81}_{-7.65}~~, &
    a_{\pi \Sigma}=&4.35^{+0.31}_{-0.77}~~, \nonumber \\
    a_{\eta \Lambda}=&-0.01^{+0.22}_{-1.07}~~, &
    a_{\eta \Sigma}=&1.90^{+0.85}_{-1.30}~~, &
    a_{K \Xi}=&15.83^{+1.85}_{-1.46}~~,
    \label{eq:uncertainty}
\end{align}
in units of $10^{-3}$. The corresponding error bands in the cross sections are shown in Fig.~\ref{Fig3}, together with the best-fit curves. It is evident that the cross sections are well constrained by the kaonic hydrogen data: there is mutual consistency between scattering and threshold measurements.

%
%

\subsection{Predictions from improved chiral SU(3) dynamics}

\subsubsection{$K^- p$ subthreshold amplitudes and structure of $\Lambda(1405)$}

\begin{figure}[htb]
\begin{minipage}[t]{7cm}
\includegraphics[width=7cm]{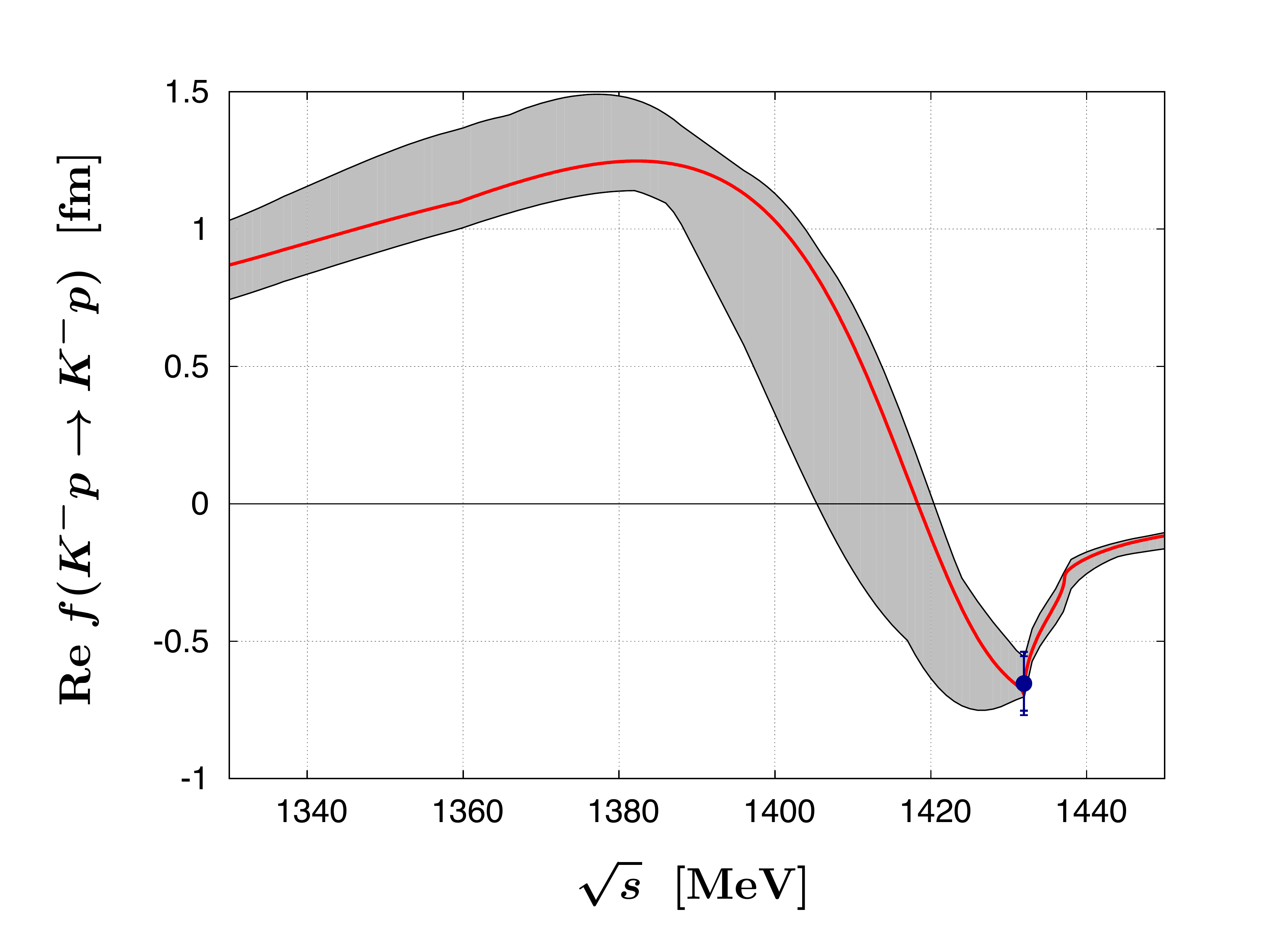}
\end{minipage}
\hspace{\fill}
\begin{minipage}[t]{7cm}
\includegraphics[width=7cm]{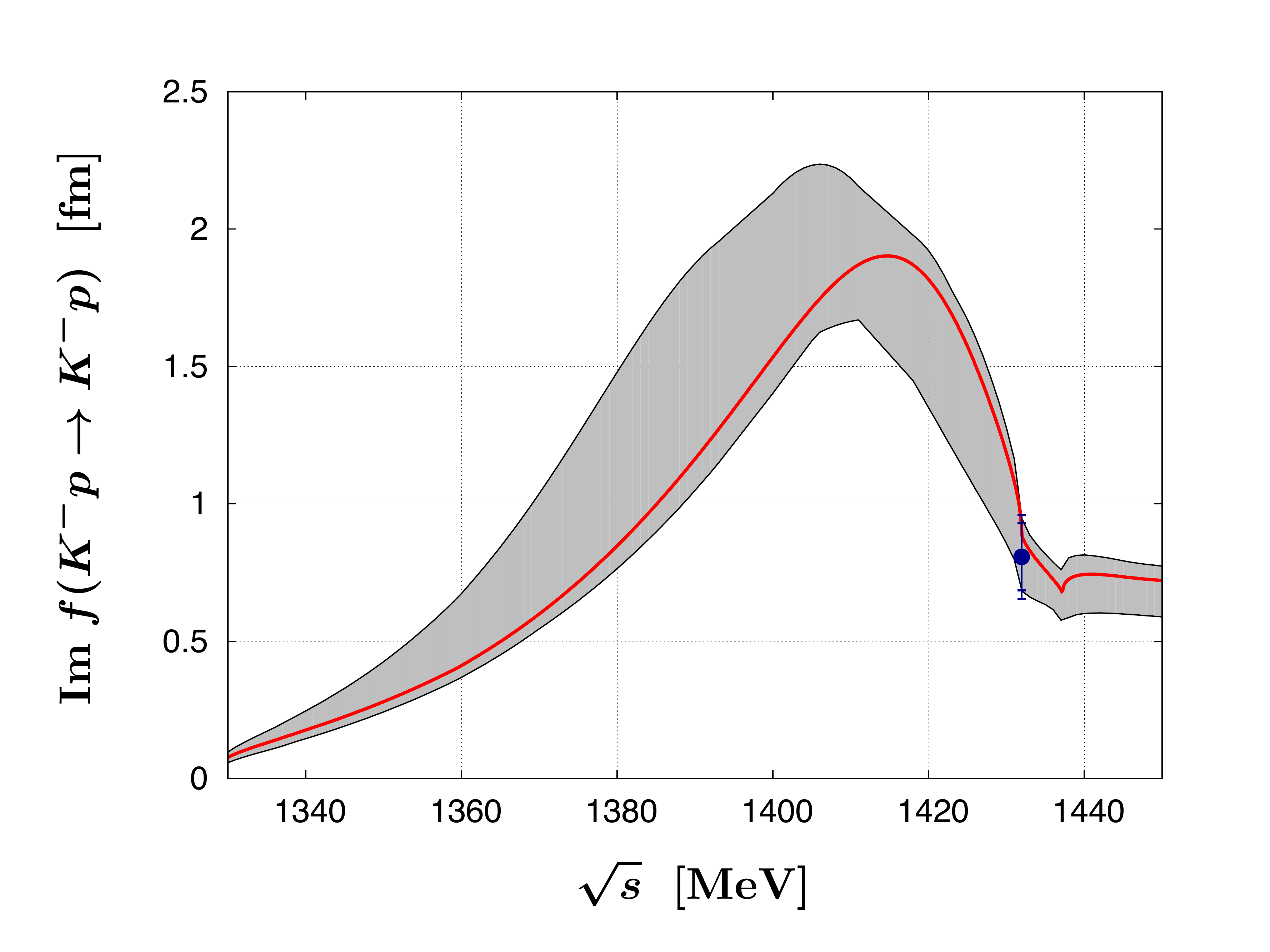}
\end{minipage}
\caption{Real part (left) and imaginary part (right) of the $K^-p \rightarrow K^-p$ forward scattering amplitude
obtained from the NLO calculation and extrapolated to the subthreshold region. 
The empirical real and imaginary parts of the $K^-p$ scattering length deduced from the recent kaonic hydrogen measurement (SIDDHARTA~\cite{Bazzi2011}) are indicated by the dots including statistical and systematic errors. The shaded uncertainty 
bands are explained in the text.} 
\label{Fig4}
\end{figure}

Once this consistent theoretical description of all $K^{-}p$ observables is obtained, it is instructive to perform extrapolations of the amplitudes to subthreshold and complex energies. This is particularly important for the understanding of the $\Lambda(1405)$ resonance as a quasibound $K^{-}p$ ($I = 0$) state embedded in the $\pi\Sigma$ continuum, as well as for the far-subthreshold $\bar{K}N$ interaction that is relevant in the context of possible $\bar{K}$-nuclear clusters. In Fig.~\ref{Fig4}, we show the subthreshold extrapolation of the real and imaginary parts of the $K^-p\rightarrow K^-p$ amplitude by the best-fit NLO scheme and the uncertainty bands constrained by Eq.~\eqref{eq:uncertainty}. As seen in the figure, the amplitude exhibits the structure of the $\Lambda(1405)$ resonance emerging from the strong attracttion in the $I=0$ component of the amplitude. We find that the subthreshold extrapolation is stable thanks to the accurate constraint at threshold.

\subsubsection{The two-poles scenario}

Next we look for poles in the second Riemann sheet of the complex energy plane to study the coupled-channels structure of the $\Lambda(1405)$ resonance. With the best-fit result in the NLO scheme, pole singularities between the $\bar{K}N$ and $\pi\Sigma$ thresholds are found at 
\begin{align}
    z_1 =& 1424 - \textrm{i} \,26 \text{ MeV}~~, &
    z_2 =& 1381 - \textrm{i} \,81 \text{ MeV}~~. \nonumber 
\end{align}
The higher energy $z_{1}$ pole is dominated by the $\bar{K}N$ channel and the lower energy $z_{2}$ pole receives stronger weight from the $\pi\Sigma$ channel. This confirms the two-poles scenario of the $\Lambda(1405)$~\cite{Oller2001,Jido2003,HW2008}. Actually, the existence of two poles around the $\Lambda(1405)$ resonance had been found in previous NLO calculations~\cite{Borasoy2005,Borasoy2006}, but the precise location of the poles, especially of the lower one, could not be determined in these earlier studies, given the lack of precision in the empirical constraints. 

In the present analysis, the SIDDHARTA measurement provides much more severe constraints also on the pole positions. The real parts of $z_1$ and $z_2$ are remarkably stable in all three TW, TWB and NLO schemes. The imaginary parts deviate within $\lesssim 20$ MeV between these schemes, as seen in Table~\ref{tab:results}. Using the error analysis from Eq.~\eqref{eq:uncertainty} together with the best-fit NLO results, one finds:
\begin{align}
    z_1 =&1424^{+7}_{-23} - \textrm{i}\, 26^{+3}_{-14} \text{ MeV}~~, &
    z_2 =& 1381^{+18}_{-6} - \textrm{i}\, 81^{+19}_{-8} \text{ MeV}~~. 
\end{align}
The uncertainties of the pole locations are thus significantly reduced from previous work, and the two-poles structure of the $\Lambda(1405)$ is now consistently established with the constraints from the precise kaonic hydrogen measurement. 
Because of isospin symmetry, the two poles are stable against variations of the I=1 subtraction constants (the ones in the $\pi\Lambda$ and $\eta \Sigma$ channels). The error assignments in the pole positions and half widths are mainly 
reflecting the uncertainties of the $\bar{K}N$ and $\pi\Sigma$ subtraction constants.
%
%

\subsubsection{$K^- p$ and $K^- n$ scattering lengths}

A discussion of low-energy $\bar{K}$-nuclear interactions requires the knowledge of both the $K^- p$ and $K^- n$ amplitudes near threshold. The complete $\bar{K}N$ threshold information involves both isospin $I=0$ and $I=1$ channels. The $K^-p$ scattering length $a(K^-p) = [a_0 + a_1]/2$ is given by the average of the $I=0$ and $I=1$ components, whereas the $K^-n$ scattering length $a(K^-n) = a_1$ is purely in $I=1$. Note that Coulomb corrections to $a(K^-p)$ and isospin breaking effects in threshold energies may be significant~\cite{MRR2004} and must be taken into account in a detailed quantitative analysis.

We first extract the scattering length $a(K^-p)$ from the SIDDHARTA measurements~\cite{Bazzi2011} using Eq.~(\ref{eq10}). The result is:
\begin{eqnarray}
\textrm{Re}\,a(K^-p) = -0.65 \pm 0.10 ~~\textrm{fm}~~,~~~\textrm{Im}\,a(K^-p) = 0.81\pm 0.15 ~~\textrm{fm}~~,
\label{scattlength}
\end{eqnarray}
where the uncertainties reflect the experimental errors. The predictions from chiral SU(3) dynamics, proceeding again through the sequence of TW, TWB and full NLO schemes, gives the following values for the $K^-p$ scattering length:
\begin{eqnarray}
a(K^-p) &=& -0.93 + \textrm{i}\, 0.82~\textrm{fm}~~\textrm{(TW)}~~,\\
a(K^-p) &=& -0.94 + \textrm{i} \,0.85~\textrm{fm}~~\textrm{(TWB)}~~,\\
a(K^-p) &=& -0.70 + \textrm{i} \,0.89~\textrm{fm}~~\textrm{(NLO)}~~.
\end{eqnarray}
The large magnitude of $\text{Re}\, a(K^-p)$ in the TW and TWB schemes corresponds to the overestimation of the kaonic hydrogen energy shift in these approaches, while the best-fit NLO result is fully compatible with the value~\eqref{scattlength} deduced from the experimental data.

To calculate the $K^{-}n$ scattering length, we construct the coupled-channels amplitudes in the charge $Q=-1$ sector ($K^{-}n$, $\pi^{-}\Lambda$, $\pi^{-}\Sigma^{0}$, $\pi^{0}\Sigma^{-}$, $\eta \Sigma^{-}$ and $K^{0}\Xi^{-}$), again using physical meson and baryon masses in order to take into account isospin breaking effects in the threshold energies. With the same subtraction constants as in the $Q=0$ sector, the calculated $K^{-}n$ scattering lengths are:
\begin{eqnarray}
a(K^-n) &=& 0.29 + \textrm{i} \,0.76~\textrm{fm}~~\textrm{(TW)}~~,\\
a(K^-n) &=& 0.27 + \textrm{i} \,0.74~\textrm{fm}~~\textrm{(TWB)}~~,\\
a(K^-n) &=& 0.57 + \textrm{i} \,0.73~\textrm{fm}~~\textrm{(NLO)}~~.
\end{eqnarray}
The relatively large jump in Re$\,a(K^-n)$ when passing from ``TW'' and ``TWB'' to the best-fit ``NLO'' scheme is
strongly correlated to the corresponding change in Re$\,a(K^-p)$. Thus, to determine the $I=1$ component of the $\bar{K}N$ scattering length, it is highly desirable to extract the $K^{-}n$ scattering length, e.g. from a precise measurement of kaonic deuterium~\cite{Meissner:2006gx,DM2011}.

\begin{figure}[htb]
\begin{minipage}[t]{7cm}
\includegraphics[width=7cm]{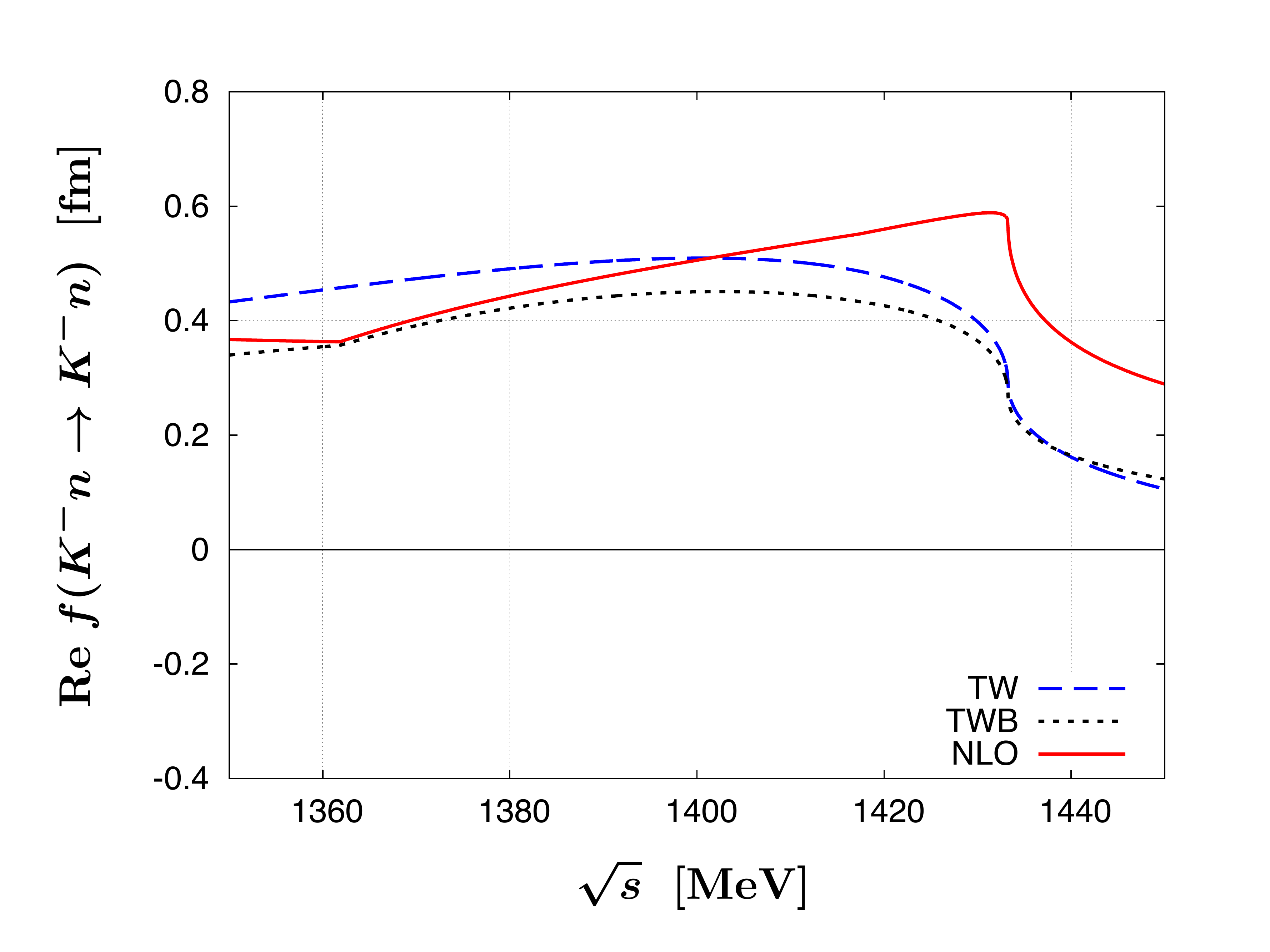}
\end{minipage}
\hspace{\fill}
\begin{minipage}[t]{7cm}
\includegraphics[width=7cm]{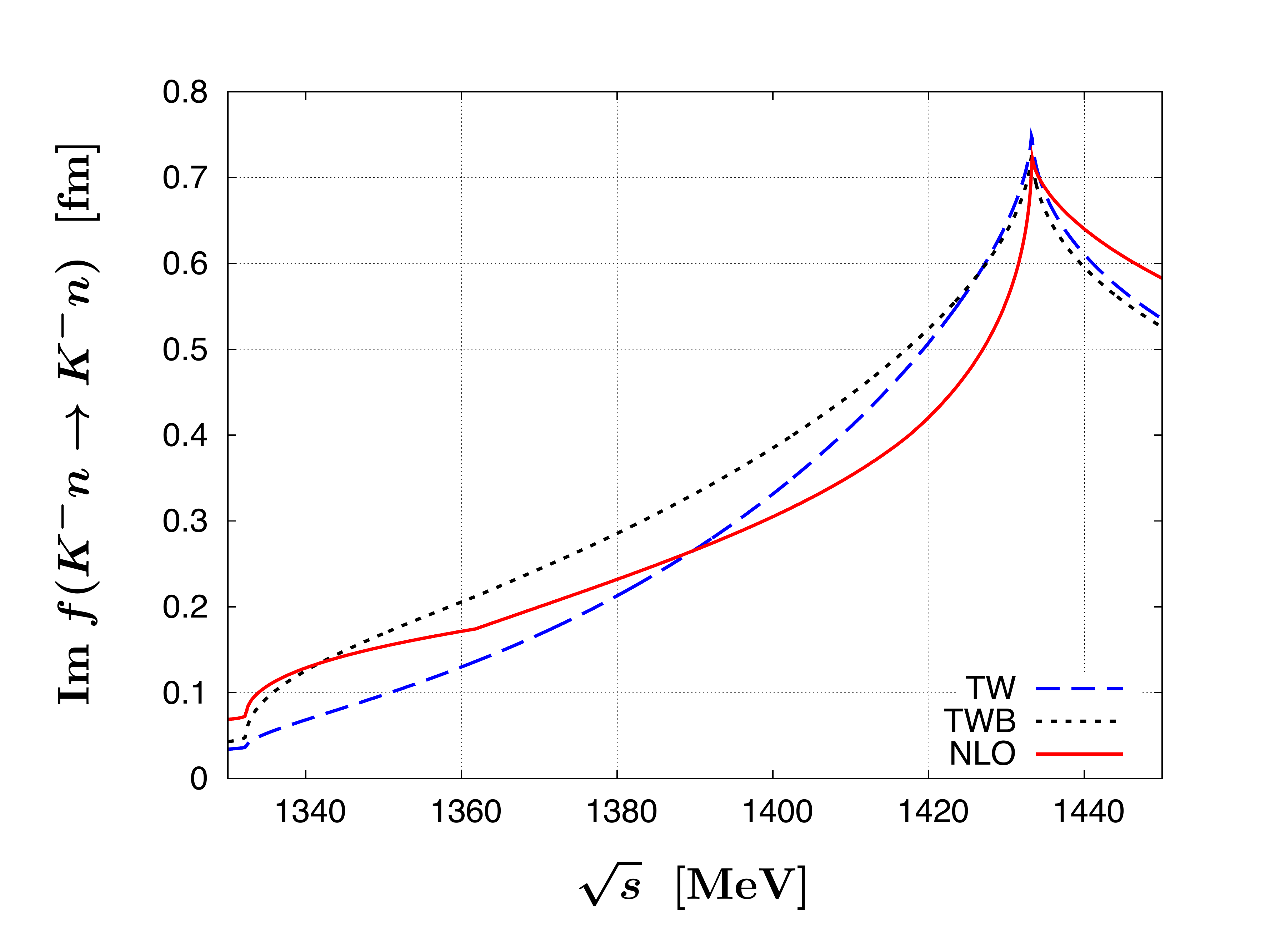}
\end{minipage}
\caption{
Real part (left) and imaginary part (right) of the $K^-n \rightarrow K^-n$ 
forward scattering amplitude extrapolated to the subthreshold region. 
} 
\label{Fig5}
\end{figure}

Next, consider the subthreshold extrapolation of the complex elastic $K^-n$ amplitude. Fig.~\ref{Fig5} shows the real and imaginary parts of this amplitude. Note that the $I=1$ $\bar{K}N$ interaction is also attractive but weaker than the $I=0$ interaction so that $f(K^-n\rightarrow K^-n)$ is non-resonant. In the absence of empirical threshold constraints for the $K^-n$ scattering length one still faces relatively large uncertainties. Variation of the subtraction constants within the range of Eq.~\eqref{eq:uncertainty} applied to the NLO scheme leads to the following estimated uncertainties:
\begin{eqnarray}
a(K^-n) = 0.57^{+0.04}_{-0.21}+ \textrm{i}\, 0.72^{+0.26}_{-0.41}  ~~\textrm{fm}~~.
\end{eqnarray}
The errors in $a(K^-n)$ relate primarily to the uncertainty of the subtraction constant
in the $\pi \Lambda$ channel.
%
%

\subsubsection{$\pi\Sigma$ invariant mass distribution}

\begin{figure}[htb]
\begin{minipage}[t]{7cm}
\includegraphics[width=7cm]{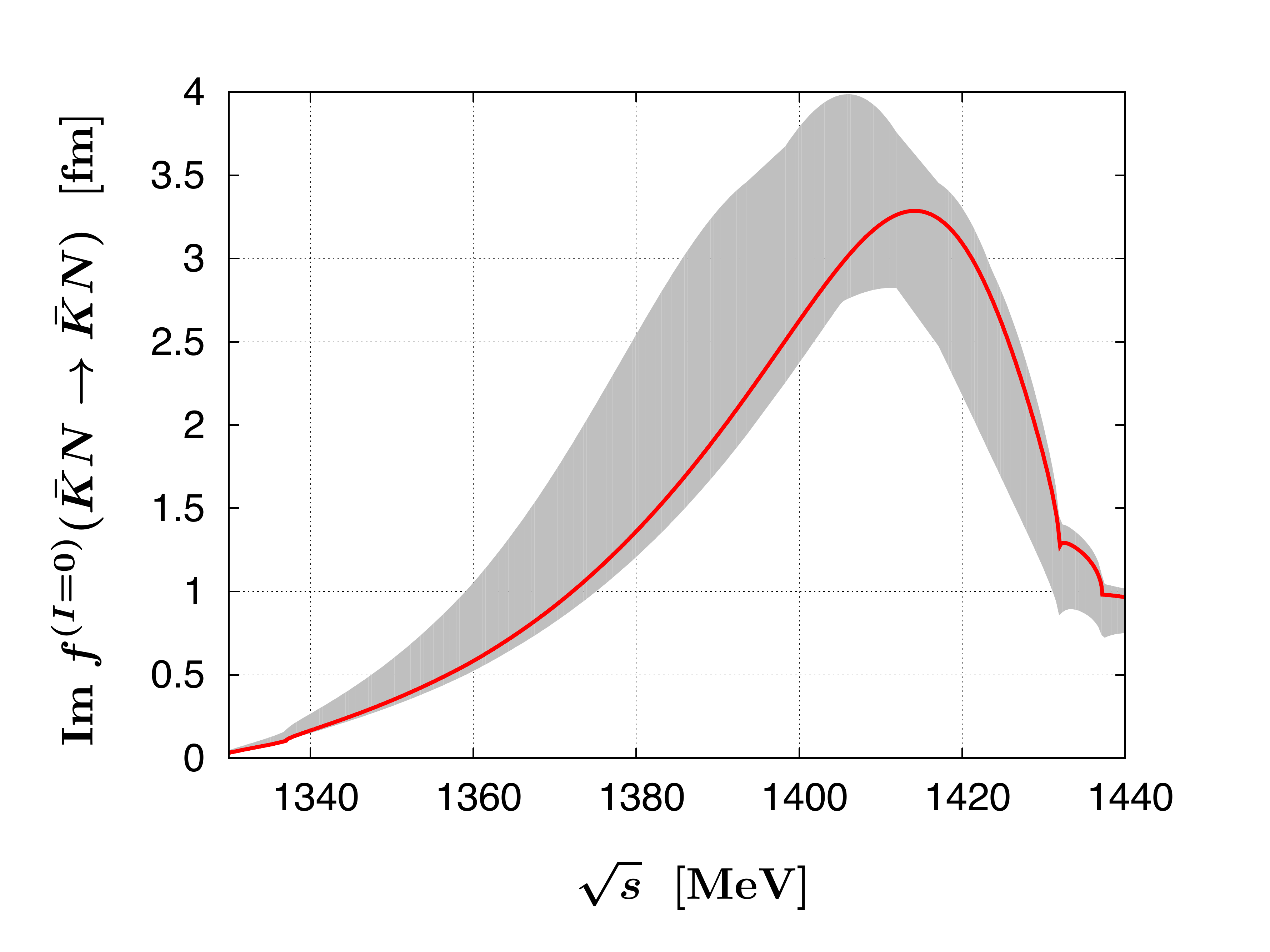}
\end{minipage}
\hspace{\fill}
\begin{minipage}[t]{7cm}
\includegraphics[width=7cm]{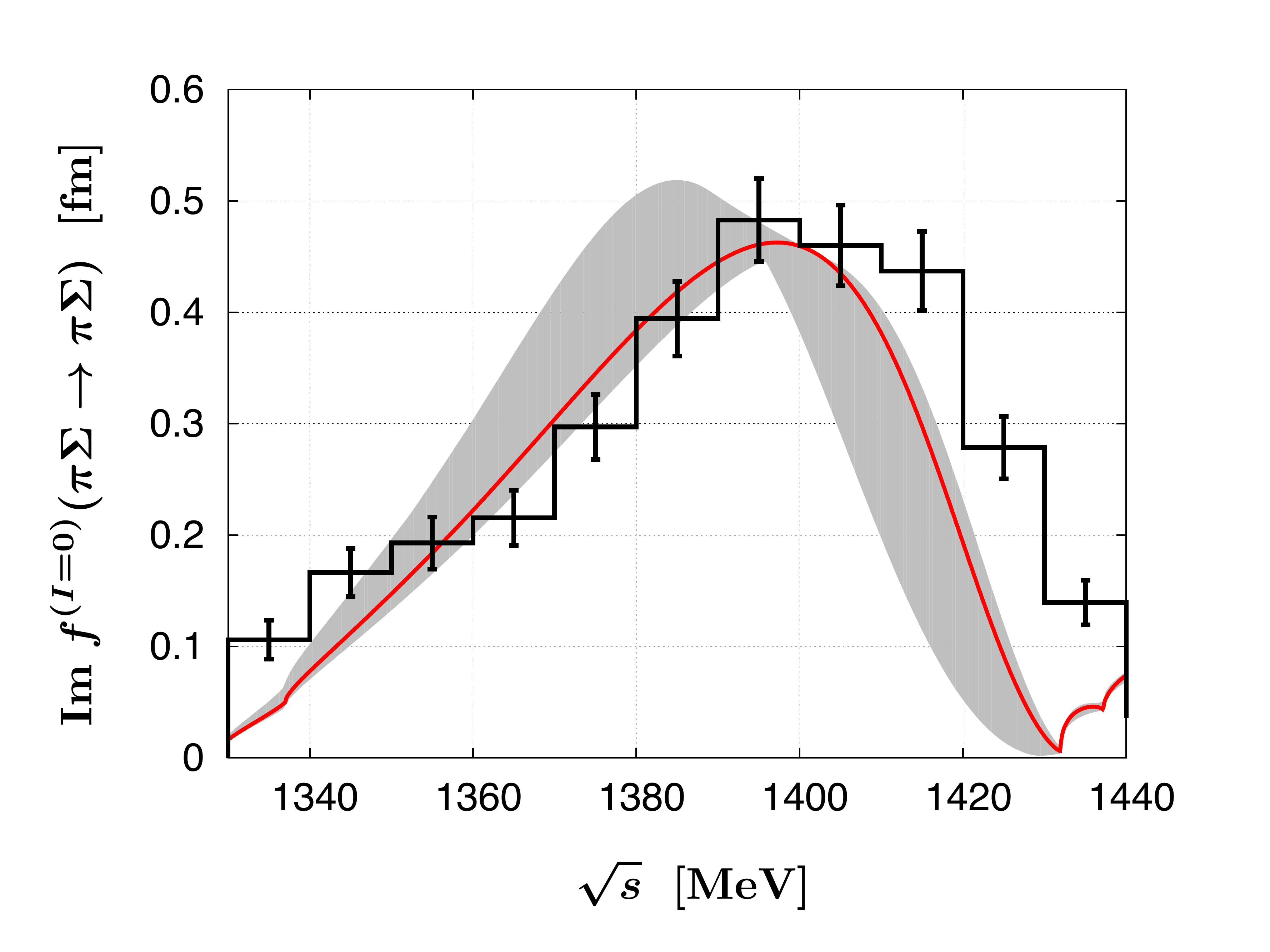}
\end{minipage}
\caption{Imaginary part of the $I=0$ $\bar{K}N$ (left) and $\pi \Sigma$ (right) amplitudes together with error bands permitted by SIDDHARTA experiments. The histogram (arbitrary unit) in the right panel denotes the experimental data of the $\pi^{-}\Sigma^{+}$ spectrum in the decay of $\Sigma^{+}(1660)\to \pi^{+}(\pi^{-}\Sigma^{+})$~\cite{Hemingway:1984pz}.
} 
\label{Fig6}
\end{figure}

One of the important features of chiral SU(3) coupled-channels dynamics is the pronounced channel dependence of the $\Lambda(1405)$ production spectra reflecting the two-poles nature of the $\Lambda(1405)$~\cite{Jido2003}. To demonstrate this two-mode structure, we show the imaginary parts of the scattering amplitudes $\pi\Sigma\to\pi\Sigma$ (Fig.~\ref{Fig6}, left) and $\bar{K}N\to\bar{K}N$ (Fig.~\ref{Fig6}, right) in the $I=0$ channel. These strength functions exhibit the $\Lambda(1405)$ spectrum as seen in different channels. Evidently, there is no single universal invariant mass distribution of the $\Lambda(1405)$. As seen in the figure, the imaginary part of the $\bar{K}N$ amplitude has its maximum close to $1420$ MeV,  whereas the position of the peak in the $\pi\Sigma$ spectrum is shifted downward from the $\bar{K}N \to \bar{K}N$ amplitude to about $1380 - 1400$ MeV. This is a consequence of the strong $\bar{K}N\leftrightarrow\pi\Sigma$ coupled-channels dynamics dictated by chiral SU(3) symmetry. The different shapes and positions of the spectral distributions in Fig.~\ref{Fig6} represent the coupled modes associated with the two poles $z_{1,2}$ discussed earlier. While the subthreshold $\bar{K}N$ spectrum has its maximum closer to the location of the ``upper" pole $z_1$, the $\pi\Sigma$ spectrum receives a stronger weight from the second, ``lower" pole $z_2$.

The right panel of Fig.~\ref{Fig6} includes for reference and orientation the experimental spectrum of the $\pi^{-}\Sigma^{+}$ channel in the decay $\Sigma^{+}(1660)\to \pi^{+}(\pi^{-}\Sigma^{+})$~\cite{Hemingway:1984pz}. It should however be noted that a direct comparison of this histogram with the imaginary part of the calculated $I=0$ $\pi\Sigma$ amplitude is not meaningful. The measured spectrum is not pure $I=0$ and the relative weights of the initial states ($\pi\Sigma$, $\bar{K}N$, \dots) are not known. In addition, because the energy of the three-body $\pi^{+}(\pi^{-}\Sigma^{+})$ system is restricted to form the $\Sigma(1660)$, the higher tail of the $\pi^{-}\Sigma^{+}$ spectrum is suppressed because of the small available phase space~\cite{HJ2011}. It is therefore necessary to construct elaborate reaction models to compare the imaginary part of the $\pi\Sigma$ amplitude with experiments, including new $\pi\Sigma$ spectra recently reported from different experiments~\cite{L1405spectrum}.

%
%

\subsection{A schematic $\bar{K}N$-$\pi\Sigma$-$\pi\Lambda$ model}

For applications in studies of strange dibaryons using sophisticated few-body techniques, a tractable model 
that starts just from the dominant leading order Tomozawa-Weinberg interaction terms in a reduced model space of $\bar KN$-$\pi \Sigma$-$\pi \Lambda$ channels, is often quite useful. The only free parameters in such a schematic model are the three subtraction constants, $a_{\bar KN}, a_{\pi \Sigma}$ and $a_{\pi \Lambda}$, in those three channels. Our aim here is to construct an ``effective" Tomozawa-Weinberg (ETW) model that reproduces the results obtained with the best-fit, full coupled-channels NLO scheme as well as possible.

\begin{figure}[htb]
\begin{minipage}[t]{7cm}
\includegraphics[width=6cm]{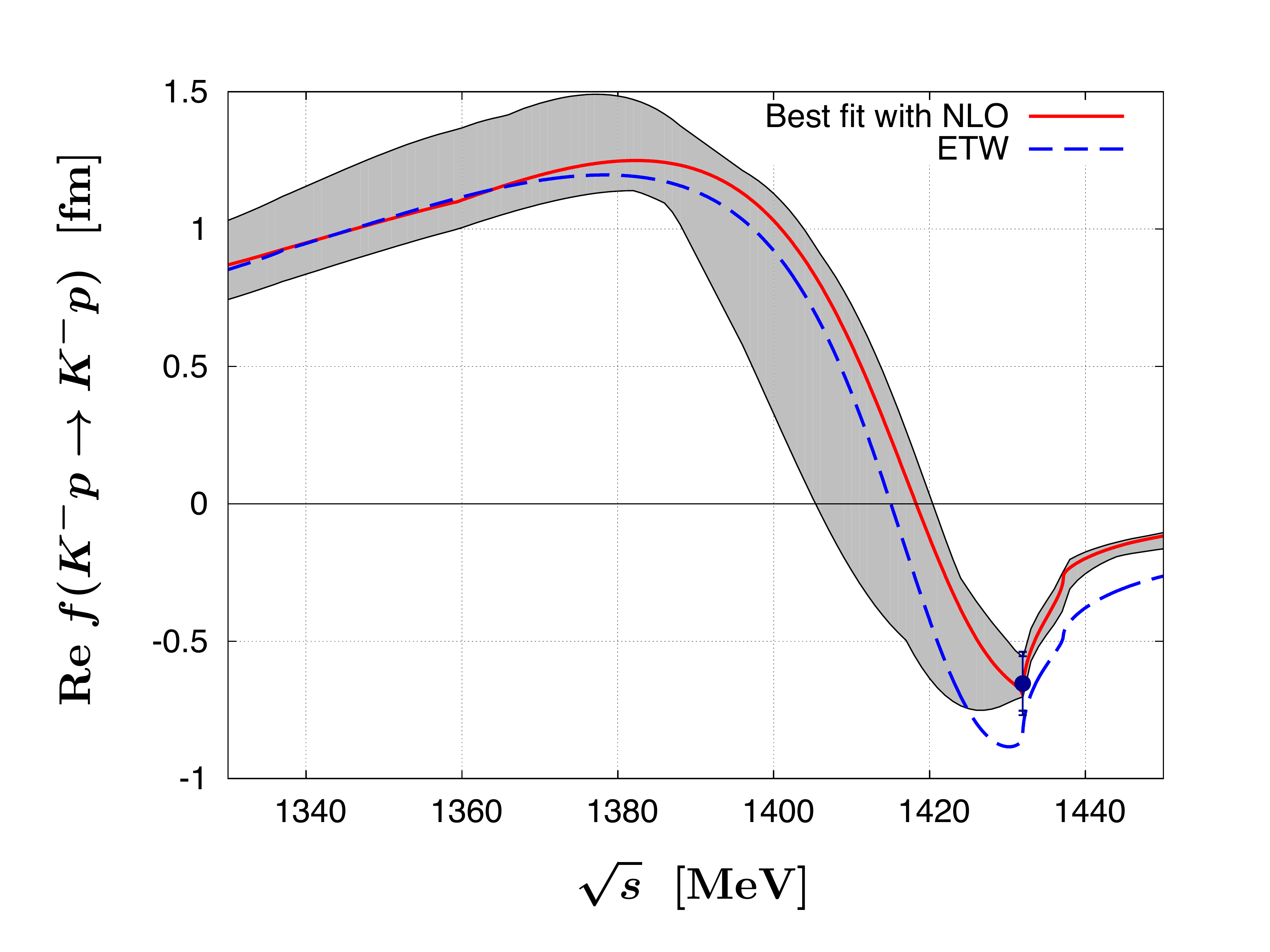}
\end{minipage}
\hspace{\fill}
\begin{minipage}[t]{7cm}
\includegraphics[width=6.2cm]{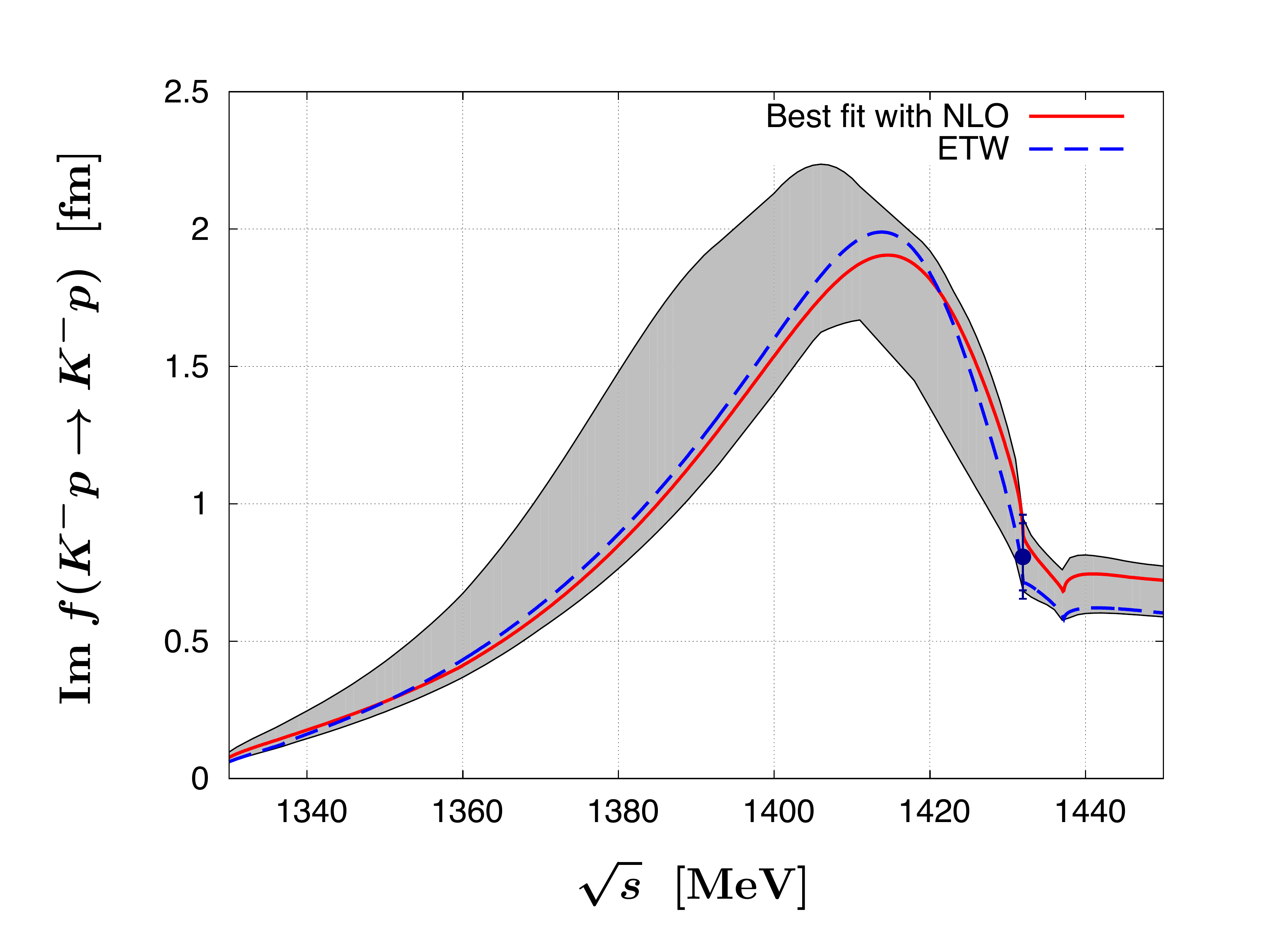}
\end{minipage}
\caption{Real part (left) and imaginary part (right) of the 
$K^-p \rightarrow K^- p$ forward scattering amplitude
together with best-fit results and with error bands permitted by the SIDDHARTA measurements. 
The dashed and solid curves denote the results of ``ETW'' and best-fit NLO, respectively.
} 
\label{Fig7}
\end{figure}

With ``natural''-sized isospin-symmetric subtraction constants, $a_{\bar KN} = -1.79\times 10^{-3}$, $a_{\pi \Sigma} = 1.81\times 10^{-3}$ and $a_{\pi \Lambda} = 7.84\times 10^{-3}$,  and with the meson decay constants $f_\pi = 92.4$ MeV and $f_{K}=109.0$ MeV, one can indeed produce a reasonable set of $\bar{K}$ threshold quantities as listed in Table~\ref{tab:ETW}. The double-pole nature of the coupled-channels dynamics is fully maintained in this schematic model. The subthreshold $K^- p \to K^- p$ amplitude is shown in Fig.~\ref{Fig7}, in comparison with the best-fit NLO results and their error bands. Although some deviations are observed around the $\bar{K}N$ threshold and the $\chi^2$ cannot compete with the one achieved in the best-fit NLO approach, the amplitudes of this ETW model for $\sqrt{s}<1425$ MeV are well within the uncertainties permitted by SIDDHARTA. They compare well with the best-fit results in the subthreshold energy region. This simplified ETW model can therefore be adopted as input in various practical applications. 
\begin{table}[tb]
  \begin{center}
    \begin{tabular}{l|l|l|l|l|l|l}  
               & $\Delta E$ [eV] & $\Gamma$ [eV] 
               & ~\,$\gamma$ & ~$R_{n}$ & ~$R_{c}$  & ~~~pole positions [MeV]\\ 
      \hline 
       ETW & ~~~338 & ~~442 & 2.26 & 0.25 & 0.62 & $1423 - 22\,i$ ~~~~$1375 - 65\,i$
    \end{tabular}
    \caption{
    Results from the effective three-channel Tomozawa-Weinberg (``ETW'') model. 
    Shown are the calculated 1s energy shift and width of kaonic hydrogen ($\Delta E$ and $\Gamma$), 
    threshold branching ratios ($\gamma$, $R_{n}$ and $R_{c}$), 
    and the pole positions of the isospin $I=0$ amplitude in the $\bar{K}N$-$\pi\Sigma$ domains.}
    \label{tab:ETW}
  \end{center}
\end{table}
%
%

\section{Summary}

We have demonstrated that the new kaonic hydrogen measurements, together with total cross section data and threshold branching ratios, are successfully described in the framework of chiral SU(3) coupled-channels dynamics with input based on the NLO meson-baryon effective Lagrangian. Our systematic study uses physical hadron masses and {\it physical} pseudoscalar meson decay constants. It reveals the hierarchy of the interaction kernel derived in chiral perturbation theory. 
It is important to point out again that the best fit to all existing data 
has been performed with the constraint that the NLO parameters of 
the chiral SU(3) meson-baryon Lagrangian stay within `natural' limits, 
such that NLO terms remain small compared to the leading-order input. 
Alternative fits using an unrestricted parameter space would be possible, with a $\chi^2$/d.o.f comparable 
to our ``best fit". 
However, in this case the NLO parameters would turn out unacceptably large 
and an unphysical pole in the $I=1$ channel below $\bar{K}N$ threshold would appear as a consequence.
If this were the only possible option, such a scenario would imply that 
the chiral SU(3) effective field theory coupled-channels approach is inconsistent and meaningless for this purpose. 
The non-trivial observation that an optimal fit can be achieved using a consistent hierarchy of LO and NLO terms, 
together with physical values of the pseudoscalar decay constants, 
justifies our conclusion that this can indeed be called a best fit.
This refined theoretical framework for the $\bar{K}N$ interaction has several important consequences.

The stringent constraints from the accurate kaonic hydrogen measurements reduce the uncertainties in the subthreshold extrapolations of the $\bar{K}N$ amplitude significantly. This is an important step towards raising the predictive power in calculations of possible dibaryon states in the $\bar{K}NN$-$\pi\Sigma N$ three-body system. The two-poles nature of the $\Lambda(1405)$ is confirmed with considerably smaller ambiguities than in previous work. The predicted imaginay parts of the $\bar{K}N$ and $\pi\Sigma$ amplitudes in the region of the $\Lambda(1405)$ show a pronounced relative shift in their peak positions, reflecting the two-modes scenario of the $\bar{K}N-\pi\Sigma$  coupled channels. The $K^{-}n$ scattering length is predicted using the same framework. Some ambiguities in the $I=1$ component of the $\bar{K}N$ interaction still remain. At present this sector is constrained by the poorly known $K^{-}p\to\pi^{0}\Lambda$ cross section data and threshold branching ratio $R_{n}$. The determination of the $K^{-}n$ scattering length, e.g. through a measurement of kaonic deuterium, would be of great importance in order to set further constraints in the dynamics of the $\bar{K}N$-$\pi\Sigma$ system.

\section*{Acknowledgements} 
We thank Avraham Gal for helpful comments and discussions. This work has been performed under the joint research cooperation agreement between RIKEN and Technische Universit\"at M\"unchen. It
is partly supported by BMBF, GSI, the DFG Cluster of Excellence ``Origin and Structure of the Universe", 
and by the Grant-in-Aid for Scientific Research from MEXT and JSPS (Nos.\ 21840026 and 23-8687).
T.H. thanks for support from the Global Center of Excellence Program by MEXT, Japan, through the Nanoscience and Quantum Physics Project of the Tokyo Institute of Technology. 

\bibliographystyle{elsarticle-num}







\end{document}